\documentclass[a4paper,fleqn,usenatbib]{mnras}

\usepackage[T1]{fontenc}
\usepackage{ae,aecompl}

\usepackage{graphicx}	
\usepackage{amsmath}	
\usepackage{amssymb}	
\usepackage{tikz}
\usepackage{hyperref}
\usepackage{times}
\usepackage[normalem]{ulem}

\usepackage{etoolbox}
\makeatletter
\patchcmd\@combinedblfloats{\box\@outputbox}{\unvbox\@outputbox}{}{%
   \errmessage{\noexpand\@combinedblfloats could not be patched}%
}%
 \makeatother

\newcommand{\WVF}{\textsc{wvf}}
\newcommand{\SVF}{\textsc{svf}}
\newcommand{\zobov}{\textsc{zobov}}
\newcommand{\SVFtwoD}{\textsc{svf\_2d}}
\newcommand{\tunnels}{tunnels}
\newcommand{\troughs}{troughs}

\newcommand{\Euclid}{\textsc{euclid}}
\newcommand{\LSST}{\textsc{lsst}}

\newcommand{\figDir}{fig_pdf/}
\newcommand{\figwidth}{0.45\linewidth}
\newcommand{\verticaloffset}{\\[-.7cm]}

\newcommand{\reduceVerticalOffset} { \vspace{-.3cm} }

\title[Void comparison in modified gravity]{The Santiago-Harvard-Edinburgh-Durham void comparison II: unveiling the Vainshtein screening using weak lensing}

\author[]
{\parbox{\textwidth}{
	Enrique Paillas$^{1,2}$\thanks{E-mail : epaillas@astro.puc.cl},
	Marius Cautun$^{3}$,
	Baojiu Li$^{3}$,
	Yan-Chuan Cai$^{4}$,
	Nelson Padilla$^{1,2}$
	Joaqu\'in Armijo$^{1,2}$
	and Sownak Bose$^{5}$
	}
\vspace{.2cm}\\
$^{1}$ Instituto de Astrof\'isica, Pontificia Universidad Cat\'olica de Chile, Av. Vicu\~na Mackenna 4860, Santiago, Chile \\
$^{2}$ Centro de Astro-Ingenier\'ia, Pontificia Universidad Cat\'olica de Chile, Av. Vicu\~na Mackenna 4860, Santiago, Chile \\
$^{3}$ Institute of Computational Cosmology, Department of Physics, Durham University, South Road, Durham, DH1 3LE, UK \\
$^{4}$ Institute for Astronomy, University of Edinburgh, Royal Observatory, Edinburgh EH9 3HJ, UK \\
$^{5}$ Harvard-Smithsonian Center for Astrophysics, 60 Garden Street, Cambridge, Massachusetts 02138, USA
}

\pubyear{2018}

\begin{document}
\label{firstpage}
\pagerange{\pageref{firstpage}--\pageref{lastpage}}
\maketitle

\begin{abstract}
    We study cosmic voids in the normal-branch Dvali-Gabadadze-Porrati (nDGP) braneworld models, which are representative of a class of modified gravity theories where deviations from General Relativity are usually hidden by the Vainshtein screening in high-density environments. This screening is less efficient away from these environments, which makes voids ideally suited for testing this class of models. We use N-body simulations of $\Lambda$-cold dark matter ($\Lambda$CDM) and nGDP universes, where dark matter haloes are populated with mock galaxies that mimic the clustering and number densities of the \textsc{boss cmass} galaxy sample. We measure the force, density and weak lensing profiles around voids identified with six different algorithms. Compared to $\Lambda$CDM, voids in nDGP are more under-dense due to the action of the fifth force that arises in these models, which leads to a faster evacuation of matter from voids. This leaves an imprint on the weak lensing tangential shear profile around nDGP voids, an effect that is particularly strong for 2D underdensities that are identified in the plane-of-the-sky.  We make predictions for the feasibility of distinguishing between nDGP and $\Lambda$CDM using void lensing in upcoming large-scale surveys such as \LSST{} and \Euclid{}. We compare with  the analysis of voids in chameleon gravity theories and find that the weak lensing signal for 3D voids is similar to nDGP, whereas for 2D voids the differences with $\Lambda$CDM are much stronger for the chameleon gravity case, a direct consequence of the different screening mechanisms operating in these theories.
\end{abstract}




\section{Introduction} \label{sec:introduction}


On scales of Megaparsecs, the Universe exhibits striking large-scale features, which are thought to have originated from random quantum fluctuations in the early Universe \citep{Zeldovich1970}. These small-scale fluctuations grew due to gravitational instabilities, and as the Universe evolved, assembled into a configuration that we refer to as the cosmic web (\citealt{Davis1985, Kirshner1987, White1987, Bond1996}). This configuration is comprised of multiple components. The highest peaks of the matter density field correspond to the nodes of the web. These nodes are interconnected by filamentary and wall-like structures of relatively moderate densities, which in turn surround very vast and under-dense regions called \textit{cosmic voids} \citep{Sheth2003,Padilla2005}.

Void regions are a key component of the large-scale distribution of matter as they account for more than 80 per cent of the total volume of the Universe \citep{Padilla2005,Cautun2013}, making them the subject of numerous studies. For example, the formation history of voids is closely related to the evolution of the other cosmic web components, since all web elements are interconnected \citep{Cautun2014a}. There appears to be a rich exchange of matter between filaments, walls and voids (e.g. \citealt{Haider2016,Paillas2017}), which could have important consequences for the physics of galaxy formation. Other works have focused on the properties of galaxies living in voids, suggesting that void galaxies could be less massive, bluer, less metal-rich and more star forming than galaxies located in denser environments (\citealt{Liu2015, Beygu2016, Beygu2017}).

Another important aspect that has gained increasing attention is the understanding that void properties are sensitive to cosmology. In particular, previous works have shown that voids respond to modifications to the underlying theory of gravity (\citealt{Li2011,Clampitt2013,Cai2014, Zivick2015,Achitouv2016}), the dark energy equation of state (\citealt{Bos2012,Pisani2015,Demchenko2016}), as well as to other alternative cosmological scenarios, such as massive neutrino cosmologies (\citealt{Villaescusa2013, Barreira2015,Massara2015,Banerjee2016, Kreisch2018}). $\rm{\Lambda CDM}$ is currently the standard cosmological model, as it is the simplest model that can account for a wide range of observed properties of the Cosmos, such as the large-scale distribution of galaxies (\citealt{Cole2005,Eisenstein2005}), the expansion history of the Universe (\citealt{Riess1998,Perlmutter1999}) and the Cosmic Microwave Background (\citealt{ODwyer2004,Hinshaw2013,Planck2016}). Despite its successes, the physics of its main ingredients remain unclear: $\rm{\Lambda}$, responsible for the late-time accelerated expansion of the Universe, and the nature of the dark matter. This has led to increased interest in alternative cosmological models that can provide an explanation for these and other unsolved problems in theoretical physics, with modified gravity being one of these models.

Models that modify the standard theory of gravity usually do so by introducing an extra scalar field, $\varphi$, which mediates a \textit{fifth force}. Since the models still need to pass very tight constraints that come from gravity tests within our Solar System, they often require screening mechanisms that mask any potential differences with respect to General Relativity (GR) on small scales \citep{Brax2013}. The nature of this screening depends on the scalar field interactions. Types of screening include the chameleon \citep{Khoury2004}, symmetron \citep{Hinterbichler2010}, dilaton \citep{Brax2010}, Vainshtein \citep{Vainshtein1972} and K-mouflage mechanisms {\citep{Babichev2009,Brax:2014wla,Brax:2014yla}}. 
Usually, the scalar field $\varphi$ can be regarded as the potential of the fifth force. The chameleon screening, for example, happens in regions of deep Newtonian potential (such as in our Solar System) where $\varphi$ becomes small, and so does the gradient of $\varphi$. The Vainshtein screening comes instead from 
derivative interactions of $\varphi$ in the scalar field equation of motion, and it suppresses the gradient of $\varphi$ rather than $\varphi$ itself. In both cases, the screening is less effective in regions with low matter density, a property that makes cosmic voids especially powerful for exploring these models and detecting potential differences with respect to GR.

In recent years, several studies have focused on using voids to constrain modified gravity theories. These include the analysis of voids in chameleon \citep[e.g.,][]{Li2011,Clampitt2013,Lam:2014kua,Cai2014,Zivick2015,Cautun2017}, Vainshtein \citep{Falck2017}, Galileon and Nonlocal gravity models (\citealt{Barreira2015,Barreira:2016ias,Baker:2018mnu}). These and other results suggest that voids are auspicious for disentangling different cosmological models, since, due to their underdense nature, the fifth force in voids is less screened than in dark matter haloes or filaments (\citealt{Bloomfield2015, Falck2015}). This fifth force can modify the void density and dynamical profiles, and in turn, the weak gravitational lensing signal that is measured around them \citep{Cai2014,Cautun2017}.

Another important aspect of void studies comes from the fact that there is a wide range of void finding algorithms in the literature \citep[see, e.g.,][]{Colberg2008,Cautun2017}, ranging from void finders that search for spherically underdense regions \citep[e.g.,][]{Padilla2005,Li2011}, to others that identify irregular voids by applying the watershed transform on a tracer density field \citep[][]{Platen2007, Neyrinck2008}. Each void finder can result in different void catalogues with different properties. This provides opportunities to identify the void finder that maximises the difference between $\rm{\Lambda CDM}$ and alternative cosmological models. In general, the optimal void finder will depend on the particular model under investigation and on the void statistics used as a test.       

This is the second in a series of papers whose goal is to improve our understanding of voids as a test of cosmological models. In this paper, we focus on a particular example of modified gravity models -- the Dvali-Gabadadze-Porrati braneworld models \citep[][DGP]{Dvali2000}, which have been proposed as an explanation for the accelerated expansion of the Universe. These models are representative of a wider class of modified gravity theories in which the screening (i.e., the recovery of GR in regions of high density) is realised by the Vainshtein mechanism. Different Vainshtein theories result in phenomenologically similar departures from GR and thus our analysis of voids in DGP gravity can be used to assess the potential of voids to test this wider class of models. By employing a variety of void finding algorithms, we aim to provide a clear picture of the void detection method that is most suited to exploring Vainshtein screening mechanism models. This is achieved by using different void statistics and by highlighting their respective advantages and disadvantages. Throughout this work we compare our results with the first paper in the series, \citet{Cautun2017}, which studied voids as a test of chameleon gravity models. This comparison allows for a detailed analysis of the impact of different screening mechanisms on the statistics of voids.

An important feature of this work is that the voids are identified in mock galaxy distributions, which are generated by applying the Halo Occupation Distribution (HOD) formalism to a set of large volume dark matter only simulations. The HOD model parameters in both the GR and nDGP halo catalogues are independently tuned to reproduce the clustering of galaxies observed in the SDSS CMASS galaxy survey \citep{Manera2013}, through which we hope that any observed differences in void statistics between the fiducial GR model and the nDGP models are not due to discrepancies in the two-point galaxy clustering. This also highlights that the modified gravity models considered here are consistent with current two-point clustering measurements, and thus motivates the necessity of finding alternative ways to constrain these theories.

The layout of the paper is as follows: In Sec. \ref{sec:Theory} we provide a brief description of the cosmological models, simulations and void finding algorithms that are used in this work. In Sec. \ref{sec:void_finders} we describe the void finding algorithms. In Sec. \ref{sec:Void profiles} we show void density, force and weak lensing profiles. Sec. \ref{sec:predictions_surveys} presents predictions for forthcoming large-scale surveys. Finally, in Sec. \ref{sec:conclusions} we list the main conclusions of the paper.

\section{Theory} \label{sec:Theory}

\subsection{nDGP cosmology} \label{subsec:ndgp}

In braneworld cosmology, the spacetime we experience is a 4D brane 
embedded in a higher-dimensional spacetime called the bulk. While in these models fundamental matter particles are usually assumed to be confined to the brane, gravitons can move through the extra spatial dimensions. This leak of gravitons to the extra dimensions provides a possible explanation as to why the strength of gravity is much weaker compared to those of the other fundamental forces \citep{Maartens2010}.

In this work we consider the normal branch of the 5D Dvali-Gabadadze-Porrati \citep[][hereafter nDGP]{Dvali2000} braneworld model, which is representative of a class of modified gravity models that feature the Vainshtein screening mechanism \citep{Vainshtein1972}. This model can have the same expansion history as $\rm{\Lambda CDM}$, which is achieved by having 
an additional dynamical dark energy component with a appropriately-tuned equation of state. On the other hand, the gravity felt by massive particles is modified due to an extra fifth force, which is mediated by a scalar field -- the brane-bending mode. The modified gravitational force is governed by one extra parameter -- the cross-over scale $r_c$ -- which delineates the transition scale at which gravity changes from 5D to 4D.

The gravitational part of the nDGP action can be written as \citep{Maartens2010}:
\begin{align} \label{eq:action_1}
S = \int_{\rm{bulk}} {\rm d}^5x\ \sqrt{-g^{(5)}} \frac{R^{(5)}}{16 \pi G^{(5)}} + 
\int_{\rm{brane}} {\rm d}^4x\ \sqrt{-g} \frac{R}{16 \pi G},
\end{align}
where $g^{(5)}$ and $g$ correspond to the determinants of the metrics of the bulk and the brane, respectively, and $R^{(5)}$ and $R$ are their associated Ricci scalars. $G^{(5)}$ and $G$ denote the gravitational constant in 5D and 4D, which are related to each other through the cross-over scale $r_c$ (which has the dimension of length):
\begin{equation}
r_c = \frac{1}{2} \frac{G^{(5)}}{G}.
\end{equation}
For scales much larger than the cross-over scale ($r \gg r_c$), gravity leaks off the brane as the first term in Eq.~\eqref{eq:action_1} dominates over the second term.

The expansion rate of this model is given by
\begin{align} \label{eq:expansion_history}
H(a) = H_0 \sqrt{\Omega_{m0}a^{-3} + \Omega_{rc} + \Omega_{\rm de0}\rho_{\rm de}(a)/\rho_{\rm de0}} - \sqrt{\Omega_{rc}} ,
\end{align}

where $H_0$ is the present-day value of the Hubble expansion rate, $a$ is the scale factor, $\Omega_{m0}$ is the present-day fractional matter density, $\rho_{\rm de}(a)$ is a dynamical dark energy density, and $\Omega_{rc}= 1 / (4 H_0^2 r_c^2)$ is a dimensionless parameter to be used in place of $r_c$.

As mentioned above, we focus on the normal branch of the DGP model (nDGP), which does not predict an accelerated expansion unless an additional dark energy component is included in the matter sector, i.e., $\rho_{\rm de}(a)\neq0$; if on the other hand $\rho_{\rm de}(a)=0$, it can be shown using Eq.~\eqref{eq:expansion_history} that $a(t)$ always decelerates. If the additional dark energy is a cosmological constant, then the expansion history would be different from that of $\Lambda$CDM; to have a $\Lambda$CDM expansion history, one has to tune the dark energy equation of state accordingly (in this work we implicitly assume that this tuning has been done so that we can take the expansion history to be $\Lambda$CDM; we also assume that the additional dark energy component does not cluster significantly so that its impact on the perturbation evolution can be neglected). The DGP model does have a self-acceleration branch, sDGP, but it is plagued by ghost problems and its predictions are at odds with cosmological observations \citep{Fang:2008kc}. Hence, strictly speaking, the DGP model does not offer a viable alternative to $\Lambda$CDM to explain the cosmic acceleration. However, as a representative example of the Vainshtein screening mechanism, it is a useful phenomenological model to study the behaviour of the Vainsthein class of models.

For massive particles, the modified Poisson equation can be written as \citep{Koyama2007}:
\begin{align}
\nabla^2 \Psi = \nabla^2 \Psi_{\rm{N}} + \frac{1}{2} \nabla^2 \varphi,
\end{align}
where $\nabla^2 \Psi_{\rm{N}} = 4 \pi G a^2 \rho \delta$, $\varphi$ is the scalar field associated to the bending modes of the brane, and $\nabla$ is the spatial derivative. 

The equation of motion for $\varphi$ reads:
\begin{equation}\label{eq:3}
\nabla^2\varphi + \frac{r_c^2}{3\beta(a)a^2}\left[ (\nabla^2\varphi)^2
- \nabla^i\nabla^j\varphi\nabla_i\nabla_j\varphi \right] = \frac{8\pi G}{3\beta(a)}\delta{\bf\rho}a^2,
\end{equation}
where Einstein's summation convention is used with $i,j$ running over $1,2,3$. The time-dependent function $\beta=\beta(a)$, which governs the strength of the fifth force in the unscreened regime, is given by:
\begin{eqnarray} \label{eq:4}
\beta(a) &=& 1 + 2H r_c \left( 1 + \frac{\dot{H}}{3H^2} \right)\nonumber\\
&=& 1 + \left[\frac{\Omega_{m0}a^{-3}+\Omega_{\Lambda0}}{\Omega_{rc}}\right]^{1/2} - \frac{1}{2}\frac{\Omega_{m0}a^{-3}}{\sqrt{\Omega_{m0}a^{-3}+\Omega_{\Lambda0}}},~~~~~~
\end{eqnarray} 
where for the second step we have assumed a $\Lambda$CDM background with $\Omega_{\Lambda0}\equiv1-\Omega_{m0}$, and an overdot denotes the derivative with respect to physical time $t$. The terms in the brackets on the left-hand side of Eq.~(\ref{eq:3}) are purely nonlinear and disappear upon linearisation of the equation, which leads to a re-expression of the modified Poisson equation:
\begin{equation} \label{eq:5}
\nabla^2 \Phi = 4 \pi G a^2 \left(1 + \frac{1}{3\beta(a)}  \right) \delta \rho \ .
\end{equation}
From Eq.~(\ref{eq:4}), we note that $\beta$ is always positive, and hence Eq.~(\ref{eq:5}) shows that the growth of structures is boosted relative to $\rm{\Lambda}CDM$. 

\subsubsection{Vainshtein screening mechanism} \label{subsubsec: Vainshtein mechanism}

In a real universe, the terms in the brackets of Eq.~\eqref{eq:3} do not vanish exactly, but on large enough scales where $|\nabla^2\varphi|\ll r_c^{-2}$, the linear term in $\nabla^2\varphi$ in Eq.~\eqref{eq:3} dominates and Eq.~\eqref{eq:5} holds to a good approximation. This happens in the regime of small density fluctuations, $\delta\rho/\bar{\rho}\ll1$, or for regions far from a massive body.

In the other limit, where $|\nabla^2\varphi|\gg r_c^{-2}$, the nonlinear terms in the brackets of Eq.~\eqref{eq:3} dominate, which means
\begin{equation}
\nabla^2\varphi \ll \frac{8\pi G}{3\beta(a)}\delta{\bf\rho}a^2 \sim 4 \pi G a^2 \rho \delta \sim \nabla^2 \Psi_{\rm N} \sim \nabla^2 \Psi \ ,
\end{equation}
where $\sim$ means "of the same order of" and we have assumed that $\beta(a)\sim\mathcal{O}(1)$ (or equivalently the fifth force has a similar strength as Newtonian gravity if unscreened). In this regime, which happens near a massive body such as the Earth, the Sun or a galaxy cluster, the fifth force, {which is given by} $\nabla\varphi/2$, is much weaker than normal gravity and can be neglected, recovering the behaviour of standard GR. This is called the screened regime.

In the case of spherical symmetry, the scalar field equation of motion can be written as:
\begin{equation}
\frac{2r_c^2}{3\beta a^2}\frac{1}{r^2}\frac{{\rm d}}{{\rm d}r}\left[r\left(\frac{{\rm d}\varphi}{{\rm d}r}\right)^2\right] + \frac{1}{r^2}\frac{{\rm d}}{{\rm d}r}\left[r^2\frac{{\rm d}\varphi}{{\rm d}r}\right] = \frac{8\pi G}{3\beta}\delta\rho a^2,
\end{equation}
where $r$ is the radial coordinate ($r=0$ at the centre of the spherical symmetry), $\delta\rho=\delta\rho(r)$ and $\varphi=\varphi(r)$. For a spherical tophat density with $\delta\rho>0$ being constant at $r\leq R$ and 0 otherwise, the above equation can be integrated to give the following solutions:
\begin{equation}\label{eq:spherical_soln1}
\frac{{\rm d}\varphi}{{\rm d}r} = \frac{4}{3\beta}\frac{r^3}{r_V^3}\left[\sqrt{1+\frac{r_V^3}{r^3}}-1\right]g_{\rm N}(r)~~~~~{\rm for~}r\geq R,
\end{equation}
and
\begin{equation}\label{eq:spherical_soln2}
\frac{{\rm d}\varphi}{{\rm d}r} = \frac{4}{3\beta}\frac{R^3}{r_V^3}\left[\sqrt{1+\frac{r_V^3}{R^3}}-1\right]g_{\rm N}(r)~~~~~{\rm for~}r\leq R,
\end{equation}
where 
\begin{equation}
g_{\rm N}(r) = \frac{GM(r)}{r^2},
\end{equation}
with $M(r)=4\pi\int^r_0{\rm d}r'r'^2\delta\rho(r')$ being the mass excess (compared to a uniform background) enclosed in $r$, and $r_V$ is the so-called Vainsthein radius defined as
\begin{equation} \label{eq:vainshtein_radius}
r_V^3 \equiv \frac{8r_c^2r_S}{9\beta^2} \equiv \frac{4 G M}{9 \beta^2 H_0^2 \Omega_{rc}},
\end{equation}
with $r_S=2GM$ the Schwarzschild radius. 
The Vainshtein radius is a useful quantity when discussing Vainshtein screening. From Eq.~\eqref{eq:spherical_soln1} it can be seen that ${\rm d}\varphi/{\rm d}r\ll g_{\rm N}(r)$ for $r\ll r_V$ (screened regime), while ${\rm d}\varphi/{\rm d}r\rightarrow\frac{2}{3\beta}g_{\rm N}(r)$ for $r \gg r_V$ (unscreened regime).

The above calculation applies to both overdense and underdense regions, with the only difference being the value of the mass excess, $M$. For overdense regions, such as haloes, $M$ is positive and $r_V$ takes positive values. In contrast, for voids $M$ is negative, which means that $r_V^3$ is negative. This means that spherically symmetric underdense regions are fully unscreened. In reality, voids are not spherically symmetric and contain local mass overdensities, such as haloes. Thus, underdense regions can still be partially screened due to the presence of local overdensities. For the two models studied in this work, we have $\Omega_{rc}=0.01$ (N5; see the next section for more information) and $0.25$ (N1) respectively. As an example, Eq.~(\ref{eq:vainshtein_radius}) can be used to show that the Vainshtein radius for a dark matter halo with a mass equal to 200 times the critical density at $R_{200} = 1\ \rm{Mpc}$ at $z=0.0$ is of the order of 40 and 20 $\rm{Mpc}$ for N5 and N1, respectively.

For the Vainshtein screening  mechanism, the screening is mainly related to the total mass enclosed in a radius and has little dependence on environment \citep{Falck2017}. This is in contrast to the chameleon screening mechanism \citep{Khoury2004}, where the screening depends on whether an object is inside an overdense or underdense environment (i.e., there is a contribution of environmental screening to the overall behaviour of the fifth force).
We expect that such a crucial difference can cause different modified gravity impacts on the lensing signals at the edges of voids which are defined by galaxies and their host dark matter haloes that are massive objects. In Vainshtein models, the weaker impact of environments on the screening \citep{Platscher2018} means that gravity at overdense void edges can more easily go back to GR due to the screening of the haloes therein, even if the region is next to a large underdensity. In contrast, in chameleon models, the vicinities of haloes are less likely to be screened if they are next to a void. Therefore we expect the modified gravity effect in the Vainshtein class of models to be different from that of chameleon models.

\subsection{Simulations and galaxy catalogues}
\label{subsec:simulations_and_catalogues}

\begin{table}
    \centering
    \caption{The HOD parameters at two redshifts, $z=0.0$ and $z=0.5$, corresponding to the three cosmological models studied here: GR and the two nDGP models, N5 and N1. The GR parameters are the ones corresponding to the \textsc{boss} \textsc{cmass} \textsc{dr9} \citep{Manera2013}, while the nDGP ones correspond to the best fitting HOD that matches the number density and two-point correlation function of the GR HOD galaxy distribution.
    }
    \label{table:HOD_parameters}
\begin{tabular}{@{}llll}
\hline\hline
HOD parameter & GR & N5 & N1\\
\hline
$z = 0.0$ \\ 
\hline
$\log(M_{\rm{min}})$ & $13.090$ & $13.098$ & $13.102$ \\
$\log(M_0)$ & $13.077$ & $13.079$ & $13.086$ \\
$\log(M_1)$  & $14.000$ & $14.019$ & $14.062$ \\
$\sigma_{\log(M)}$ & $0.569$ & $0.607$ & $0.653$ \\
$\alpha$ & $1.010$ & $1.013$ & $1.013$ \\
\hline
$z = 0.5$ \\ 
\hline
$\log(M_{\rm{min}})$ & $13.090$ & $13.104$ & $13.100$ \\
$\log(M_0)$ & $13.077$ & $13.078$ & $13.076$ \\
$\log(M_1)$  & $14.000$ & $14.022$ & $14.046$ \\
$\sigma_{\log(M)}$ & $0.569$ & $0.604$ & $0.604$ \\
$\alpha$ & $1.010$ & $1.013$ & $1.013$ \\
\hline\hline
\end{tabular}
\end{table}

\begin{figure}
	\includegraphics[width=\columnwidth]{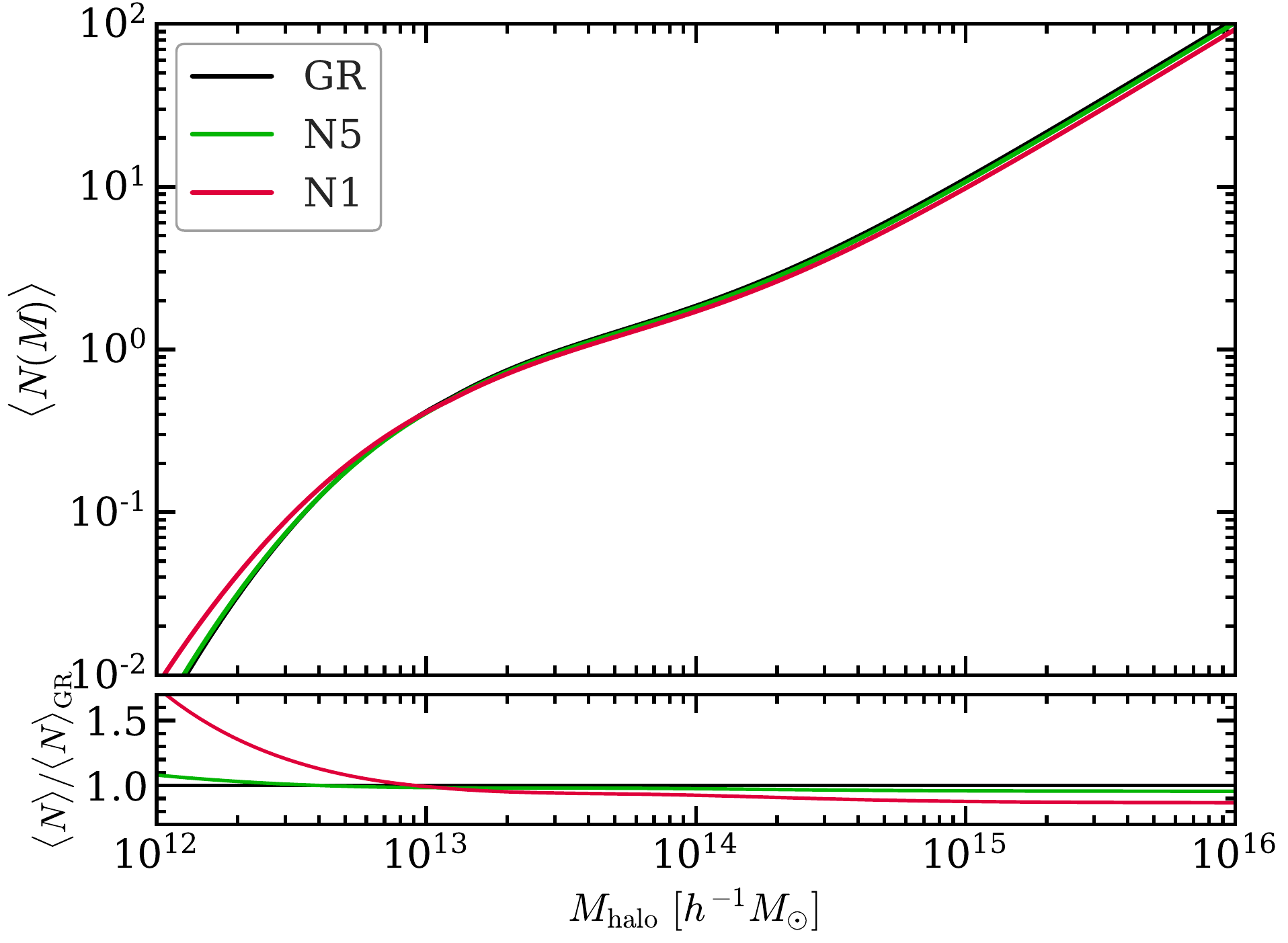}
    \vskip -.2cm
    \caption{The mean number of galaxies, $\langle N \rangle$, as a function
    of host halo mass, $M_{\rm{halo}}$, for our HOD catalogues. The different curves show the GR, N5 and N1 HOD models at redshift, $z = 0.0$. The lower panel shows the ratio between nDGP and GR.
    }
    \label{fig:HOD_nDGP_Box2}
\end{figure}

In this work we use the {\sc elephant} simulations (Extended LEnsing PHysics using ANalaytic ray Tracing; introduced in \citealt{Cautun2017}), which consist of a suite of dark-matter-only N-body simulations of the standard cosmological model, $\Lambda$CDM, and of modified gravity theories, such as nDGP and $f(R)$ chameleon gravity. These were run using the {\sc ecosmog} code \citep{ecosmog,Li:2013nua,Bose:2016wms}, an augmented version of the adaptive-mesh-refinement N-body and hydrodynamic simulation code {\sc ramses} \citep{Teyssier2002}. {\sc ecosmog} implements new subroutines to solve the additional dynamical degree of freedom (scalar field $\varphi$) present in modified gravity theories. Details about the code can be found in \citet{ecosmog,Li:2013nua}, and interested readers are referred to these manuscripts. 

Each simulation in this suite consists of a $(1024~h^{-1}\rm{Mpc})^3$ periodic volume that has been simulated using $1024^3$ dark matter particles, with a mass resolution of $7.79 \times 10^{10}\ h^{-1} M_{\rm{\odot}}$. The cosmological parameters, which are the same for both the GR and modified gravity runs, are chosen as the best-fit values from the WMAP 9-year result \citep{Hinshaw2013}: $\Omega_m = 0.281$, $\Omega_\Lambda = 0.719$, $h = 0.697$, $n_s = 0.971$ and $\sigma_8 = 0.82$. The simulations were started from $z_{\rm{init}} = 49$, with initial conditions generated using the \textsc{mpgrafic} package \citep{mpgrafic}. 

Here, we make use of the {\sc elephant} simulations for GR and two variants of the nDGP  model, where, for each model, we have simulated five independent realizations that differ only in their random phases. To minimise sample variance, each GR realization has a corresponding modified gravity simulation that shares the same initial conditions (i.e. the same random phases), but whose dynamics have been evolved using nDGP equations of motion, and not GR. The two nDGP models used here reproduce the $\rm{\Lambda CDM}$ background expansion and  differ only in the strength of their deviations from GR. The first model, called N1, has $H_0 r_c = 1$ and deviates the most from GR, while the second one, called N5, has $H_0 r_c = 5$ and is phenomenologically closer to GR. 
The values of the $H_0 r_c$ parameter combination were chosen to match the $\sigma_8$ values at $z=0$ of two $f(R)$ models: N1 corresponds to F5, which has $|f_{R_0}|=10^{-5}$, and N5 to F6, which has $|f_{R_0}|=10^{-6}$. The statistics of voids in F5 and F6 models have been studied before \citep{Cai2014,Cautun2017} and thus we can have a one-to-one comparison of the extent to which voids discriminate between $f(R)$ and nDGP modified gravity models.

Dark matter haloes and their self-bound substructures are identified using the phase-space friend-of-friend halo finder \textsc{rockstar} \citep{Behroozi2013}. We populate dark matter haloes in the $z = 0.0$ and $z = 0.5$ snapshots of the simulations using the Halo Occupation Distribution method (HOD; \citealt{Peacock2000,Scoccimarro2001,Benson2000,Berlind2002,Kravtsov2004}), which assumes that the probability that a dark matter halo hosts a certain number of galaxies correlates with the host halo mass through a simple functional form. Here we adopt the form presented in \cite{Zheng2007}:
\begin{align}
\langle N_{\rm{cen}}(M) \rangle &= \frac{1}{2} \Big[ 1 + \rm{erf}
\left( \frac{\log(M) - \log(M_{\rm{min}})}{\sigma_{\log(M)}} \right)\Big] \,\\
\langle N_{\rm{sat}}(M) \rangle &= \langle N_{\rm{cen}} \rangle
\left( \frac{M - M_0}{M_1} \right)\ ,
\end{align}
in which $\langle N_{\rm{cen}}(M) \rangle$ and $\langle N_{\rm{sat}}(M) \rangle$ are the average numbers of central and satellite galaxies inside a host halo of mass $M$, respectively. The {mean} number of galaxies in a halo of mass $M$ is then given by $\langle N_{\rm{cen}}(M) \rangle + \langle N_{\rm{sat}}(M) \rangle$. The symbols: $M_{{\rm min}}$, $M_0$, $M_1$, $\sigma_{\log(M)}$ and $\alpha$, denote free parameters of the model. For the GR mocks, we use the HOD parameters of \citet{Manera2013}, which were calibrated to create a mock galaxy catalogue representative of the \textsc{boss} \textsc{cmass} \textsc{dr9} galaxy sample. For our modified gravity simulations we tuned all five HOD parameters individually to match their galaxy number densities and two-point correlation functions with those for the corresponding realisations of the GR simulation.

The HOD parameters that were used for one of the realisations of the GR, N5 and N1 mock galaxy catalogues are given in Table \ref{table:HOD_parameters}. Fig.~\ref{fig:HOD_nDGP_Box2} shows the average number of galaxies as a function of host halo mass for these nDGP and GR HOD models at $z=0.0$. For more details about the construction of these HOD catalogues, the reader is referred to \citet{Cautun2017}. Note that the HOD occupations are qualitatively different between the nDGP models studied here and $f(R)$ models studied in \cite{Cautun2017}: low-mass haloes in nDGP tend to host more galaxies than in GR, while it is the opposite in $f(R)$ gravity, which is the reflection of the fact that $f(R)$ gravity significantly enhances the abundance of small haloes, while at $\sim10^{13}h^{-1}M_\odot$ the nDGP models studied here show almost the same halo abundance as GR.

\section{Void finders} \label{sec:void_finders}

Using the procedure described in the previous section, we populate the ELEPHANT simulations with galaxies in such a way that regardless
of gravity, they are all in agreement with CMB Planck measurements as these are adopted for their initial conditions and, at the same
time, all display the same galaxy correlation function.  We use these galaxy populations to find voids in these simulations with the aim
to find ways to still be able to detect the different theories of gravity evolved in the simulations.

We use six different void finding algorithms: three of them make use of the three-dimensional (3D) galaxy distribution to identify 3D voids, while the other three make use of the projected distribution of galaxies on the simulated mock sky to identify 2D voids. For simplicity, we neglect redshift space distortions in galaxy coordinates, and identify 3D voids in real space. For the 2D voids, we construct the projected distribution of galaxies in the distant observer approximation and project the entire simulation box (side length $1024~h^{-1}\rm{Mpc}$) along one of its principal axes. The projected simulation box corresponds to roughly the comoving distance between redshift 0.3 and 0.7. 
The following subsections provide a short summary of each of these void finders. For a more detailed description, we refer the reader to Sec. 3 in \citet{Cautun2017} and the papers introducing and testing the void finders. 

\begin{figure*}
	\includegraphics[width=\textwidth]{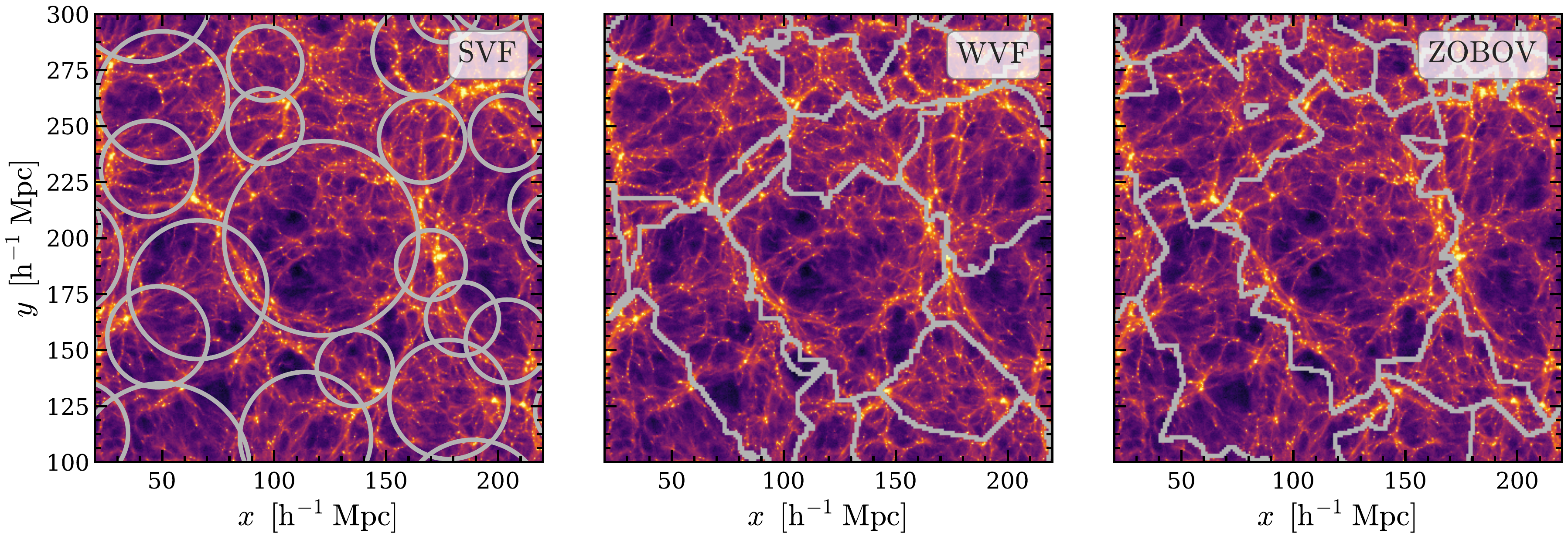}
    \vspace{-.5cm}
    \caption{A visual comparison of the 3D void finding algorithms. The grey contour lines show voids identified by the \SVF{} (left), \WVF{} (middle) and \zobov{} (right) methods in the $z=0$ snapshot of a GR simulation. The background colour map shows the dark matter density in a $50h^{-1}$Mpc thick slice along the z-axis of the simulation.
    }
    \label{fig:density_map}
    \vspace{-.1cm}
\end{figure*}

\subsubsection*{3D Spherical void finder (\SVF{})}

The \SVF{}\footnote{\href{https://github.com/epaillas/SVF}{https://github.com/epaillas/SVF}} identifies spherical under-densities in a 3D galaxy distribution. It starts by constructing a $256^3$ regular grid over the galaxy distribution, and counts the number of galaxies in each grid cell. Spheres are grown around empty grid cells, and the maximal sphere around each centre that has an integrated density of 20 per cent of the mean galaxy number density is considered as a prospective void. In order to filter voids that might have very similar centres and volumes, if two voids have centres that lie closer than 20 per cent of the sum of their radii, the smaller of them is removed from the catalogue. From the voids that are left from the previous step, we verify if the radius can increase by shifting the void centre in different directions around the original position. If the new radius is larger than the original, the position of the void is updated. A final overlapping filtering is performed, where if two voids have centres that lie closer than 80 per cent of the sum of their radii, the smaller of the two is excluded from the catalogue \citep{Padilla2005,Cai2014,Paillas2017}.

\subsubsection*{3D Watershed void finder (\WVF{})}

The \WVF{} \citep{Platen2007} identifies voids as the watershed basins of the density field. The first step consists of constructing the density on a regular grid (we use a grid spacing of $1~h^{-1}\rm{Mpc}$) by applying  Delaunay Tessellation Field Estimator \citep{Schaap2000,Cautun2011} to the galaxy distribution. Then, to remove small spurious voids, the density is smoothed with a $2~h^{-1}\rm{Mpc}$ Gaussian filter. Finally, the smoothed density is split into watershed basins, with a basin being composed of all the grid cells whose path of steepest descent (i.e., similar to the path of a rain drop along a landscape) ends at the same density minimum. Since \WVF{} voids have irregular shapes, the centre of a void is given by the volume-weighted barycentre of all the grid cell associated to that void. The void radius corresponds to that of a sphere with the same volume as the void volume.

\subsubsection*{3D \zobov{} void finder}

We use a modified version of the \zobov{} algorithm presented in \citet{Neyrinck2008}. \zobov{} estimates the galaxy density field at each galaxy position by constructing a Voronoi tesselation and compares each Voronoi cell with its neighbours to identify the density minima. Cells of increasing densities are joined together to form 'zones'. These zones stop growing when the density of the next neighbouring Voronoi cell decreases. In the original algorithm, zones are grouped together if their boundary is below a specified density threshold. In our version of the algorithm, we instead consider each zone as an individual void, as \citet{Cai2017} showed that zone merging can produce spurious voids that do not correspond to true matter under-densities. Similar to \WVF{}, the effective void radius is that of a sphere with the same volume as the void.

\subsubsection*{2D Spherical void finder (\SVFtwoD{})}

The \SVFtwoD{} works similarly to its 3D analogue, but uses the 2D galaxy distribution as the tracer field. It starts by constructing a $2048^2$ regular grid over the projected distribution of galaxies along a given axis of the simulation, and circles are grown around grid cells that are empty of galaxies. The maximal circle around each centre that has a 2D integrated density equal to 40 per cent of the mean galaxy number density is considered as a prospective void. If two voids have centres that lie closer than 80 per cent of the sum of their radii, the smaller of them is removed from the catalogue. The positions of the resulting voids are shifted around the original centres to verify if the radius can increase, and if so, the the void centre is updated. As a final step, if two voids have centres that lie closer than 20 per cent of the sum of their radii, only the larger one is kept in the catalogue.

The integrated density criterion adopted, which is different than for the \SVF{}, was calibrated to produce the strongest weak lensing detection by 2D under-densities {\citep{Cautun2017}}.

\subsubsection*{2D Tunnels}

The tunnels correspond to circular regions in the projected distribution of galaxies that are devoid of any galaxies \citep{Cautun2017}. They are identified by first constructing a 2D Delaunay tessellation using the projected galaxy distribution. By definition, the circumcircle of every Delaunay triangle is empty of galaxies, with the closest galaxies being the three located exactly on the circumcircle. The resulting catalogue is further pruned by discarding: (1) 
{all circumcircles that correspond to skinny Delaunay triangles (i.e. when the area of the triangle is less than 0.2 times that of its circumcircle)}, and (2) all circumcircles whose centre is found inside a larger circumcircle. Furthermore, we only keep the objects whose radius is $1~h^{-1}\rm{Mpc}$ or larger, which were shown in \citet{Cautun2017} to correspond to line-of-sight underdensities. The tunnel centre and radius are given by the centre and radius of its corresponding circumcircle.

\subsubsection*{2D Troughs}

Troughs are identified using the projected galaxy distribution and  correspond to fixed radius circles on the simulated sky that contain very few galaxies. They are identified by randomly placing many circles of $2~h^{-1}\rm{Mpc}$ in radius and selecting only the ones that contain 2 or fewer galaxies. This selection procedure corresponds to the 5 arcmin troughs studied by \citet{Gruen2015} and \citet{Barreira2017a}. On average, the troughs cover $23$ and $30$ per cent of the simulated sky for the $z=0$ and $z=0.5$ mock catalogues, respectively. This is much smaller than for the other {two 2D} void finders, {which cover most of the simulated sky}.

\subsubsection{{A visual comparison of void finders}}

\begin{figure}
	\includegraphics[width=\columnwidth]{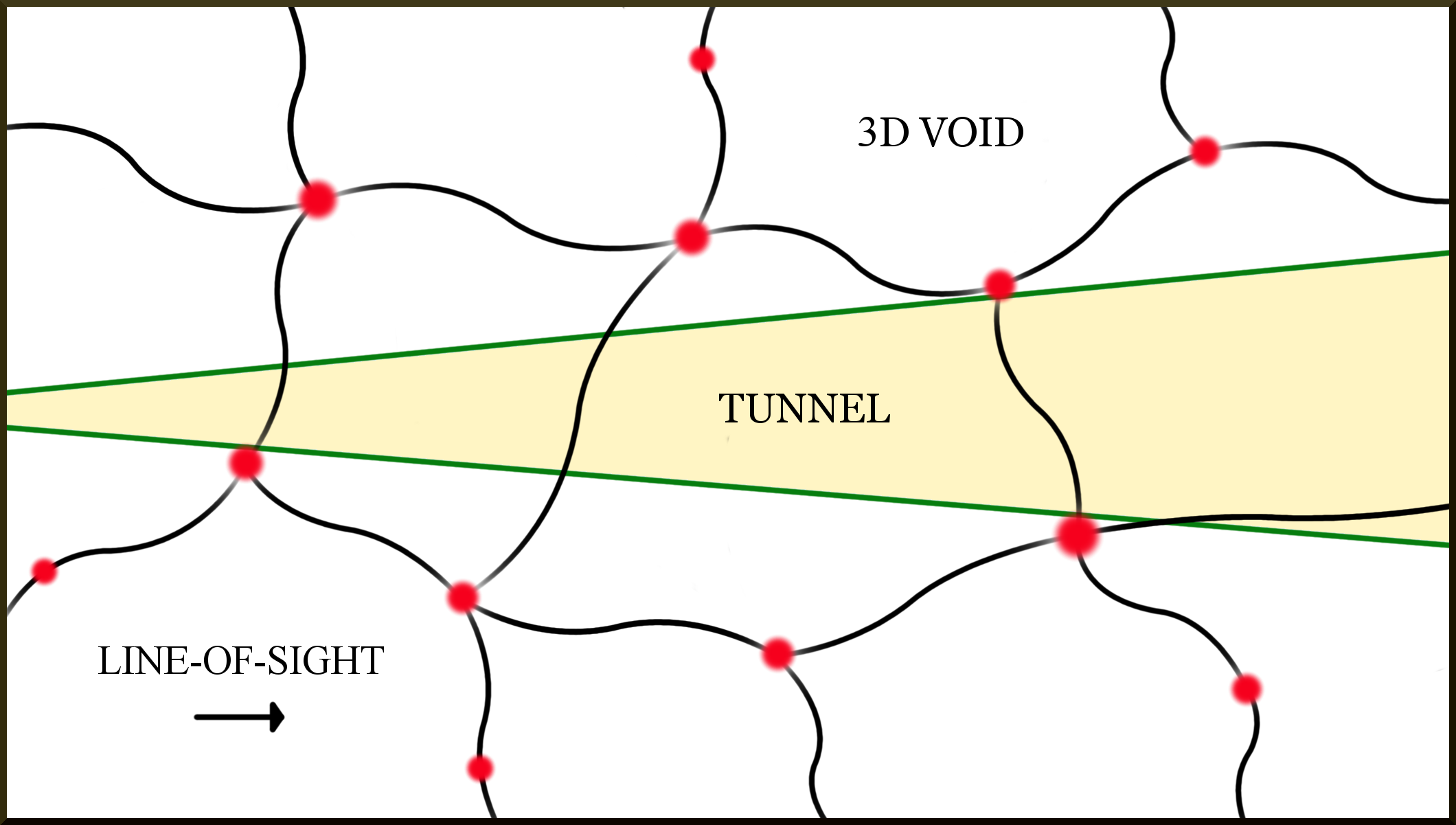}
    \vskip -.0cm
    \caption{An illustration of the geometry of 3D voids and \tunnels{}. The black lines show the boundaries of typical 3D voids, such as those identified by the \WVF{}. An example of a tunnel is shown by the yellow region that is delimited by the green lines. The tunnel is a cross section of several 3D voids along the line-of-sight, but its interior only intersects the portion of the voids boundaries that do not contain any massive haloes hosting galaxies, shown by the red dots.
     }
    \label{fig:tunnels_sketch}
\end{figure}

In Fig.~\ref{fig:density_map} we present a visual comparison between the three 3D void finding algorithms described above. Each panel shows the dark matter field in a $200 \times 200 \times 50~\left(h^{-1}{\rm Mpc}\right)^3$ sub-region of one of the GR simulations. The colour map encodes the dark matter density, where brighter colours show higher densities of dark matter. The light-grey contours show under-densities identified by the \SVF{} (left), \WVF{} (middle) and \zobov{} (right). For a visualisation of the 2D finders, the reader is referred to \citet{Cautun2017}.

We note that all 3D finders successfully identify under-dense regions of the cosmic web. While some similarities can be recognised between the \SVF{} and the finders based on space-tessellation, \WVF{}/\zobov{}, the latter are more effective in capturing the topology of the large-scale matter distribution, because these voids are not required to be spherical restricted by shape constraints as in the case of the \SVF{}. It can also be noted that some \SVF{} voids overlap with each other, as mentioned in the description of the algorithm above. This overlap allows a collection of several spherical voids to follow the non-spherical shape of the cosmic web to a better extent. The \WVF{} and \zobov{} show many similarities, as both of these algorithms apply the watershed transform to the distribution of galaxies to trace the cosmic web. However, there are differences too, with the \zobov{} method finding fewer voids than the \WVF{} one. In most cases, these differences correspond to the largest \zobov{} voids being divided into two or more voids by the \WVF{} method (e.g. the large void in the centre of Fig. \ref{fig:density_map}). This feature was highlighted when looking at the void abundance in \citet{Cautun2017}, where it was shown that, on average, \zobov{} produces fewer and larger voids than the \WVF{}.

The fact that everything in the middle and right-hand side panels of Fig. \ref{fig:density_map} seems to be part of a void is a feature of the void finding algorithms that are based on the watershed transform. As discussed in more detail in \cite{Platen2007}, the collection of watershed basins fill all the volume that is spanned by the  tracers. Each ``patch'' in Fig. \ref{fig:density_map} is considered as a single ``void'', and two adjacent ``voids'' share common boundaries, which are typically delimited by the filaments and sheets of the cosmic web. In this sense, the patches not only include the underdense interiors of the voids, but also the walls and filaments that surround them which are indeed useful when analysing the void density profiles beyond the void radius as will be evidenced in the next section.

In Fig. \ref{fig:tunnels_sketch} we illustrate the connection between 3D and 2D voids. It shows the distribution of 3D voids, such as \WVF{} ones, in a thin slice and the yellow shaded region indicates a typical tunnel. The filaments and walls of the cosmic web that define the boundaries of 3D voids are represented by the black lines, while dark matter haloes that host galaxies are shown by red dots. The interior of 2D voids corresponds to small cross-sections through several 3D voids found along the line of sight. Due to how \tunnels{} are defined, they intersect only a subset of void edges: the ones that do not contain galaxies, and hence massive haloes. On the other hand, \SVFtwoD{} voids and \troughs{} can contain a small number of galaxies.

\


\section{Void profiles} \label{sec:Void profiles}

In this section we compare void density, force and weak lensing tangential shear profiles between nDGP models and the fiducial $\rm{\Lambda CDM}$ cosmology. These results are shown for each of the void finders listed above, which were run on the HOD catalogues described in Sec.~\ref{subsec:simulations_and_catalogues}. Similar to what was found in \cite{Cautun2017} for chameleon models, there are no significant differences in the void abundance and galaxy density profiles between nDGP and $\Lambda$CDM, once the galaxy number densities and two-point galaxy clustering from each model are matched. Therefore, these results are not shown here. The redshift space distortions of the galaxy-void correlation function will be investigated in a separate paper.

Unless otherwise specified, the void profiles that are presented in this section correspond to averages of individual void profiles that were calculated over 5 different realisations of the corresponding nDGP or GR simulation. We re-scale individual profiles by the effective void radius ($R_{\rm{eff}}$) before stacking them, weighting each void by $R_{\rm{eff}}^2$. This weight is motivated by observational measurements of void weak lensing profiles: larger voids generally have more source galaxies in their background that contribute to the total lensing signal and thus their tangential shear profiles can be measured with smaller errors than for small voids. This means that by up-weighting large voids we can increase the signal-to-noise of the stacked tangential shear profile.

To estimate uncertainties, for each realisation we split the volume into $4^3$ (or $8^2$ for the case of 2D finders) non-overlapping regions and perform 100 bootstrap re-samples over these regions. Using these $5 \times 100$ samples, we compute the correlation matrix and estimate the corresponding uncertainties.

\subsection{Matter density profiles} 
\label{subsec:void_den_profiles}

\begin{figure*}
    \begin{tabular}{c}
	    \includegraphics[width=\textwidth]{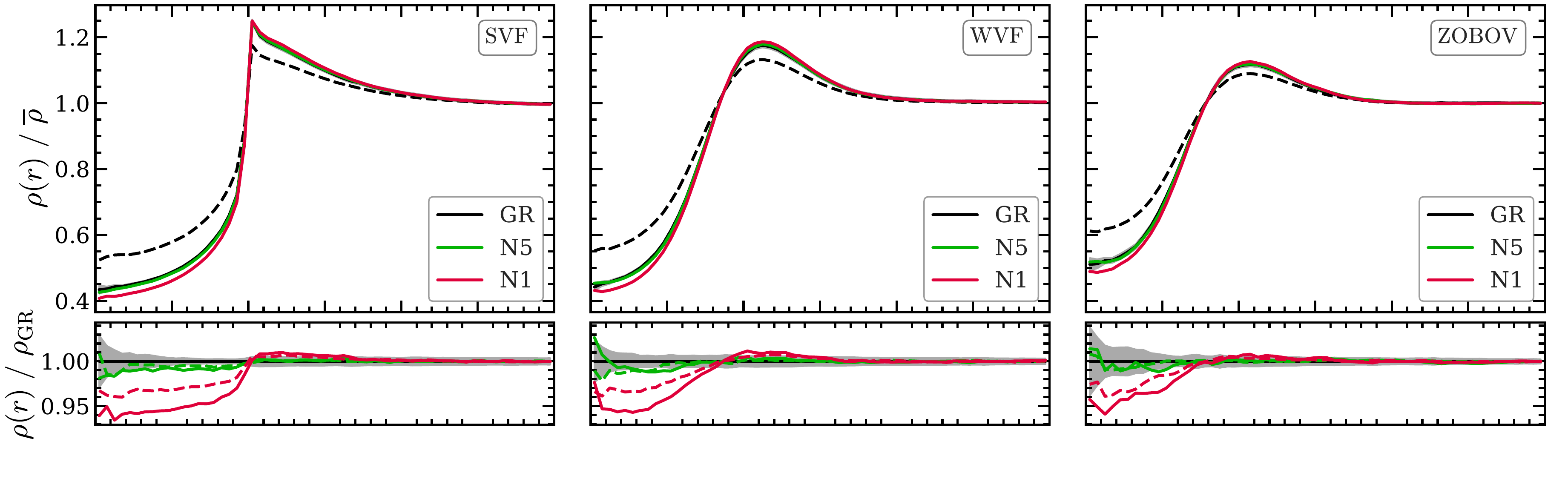} \verticaloffset
	    \includegraphics[width=\textwidth]{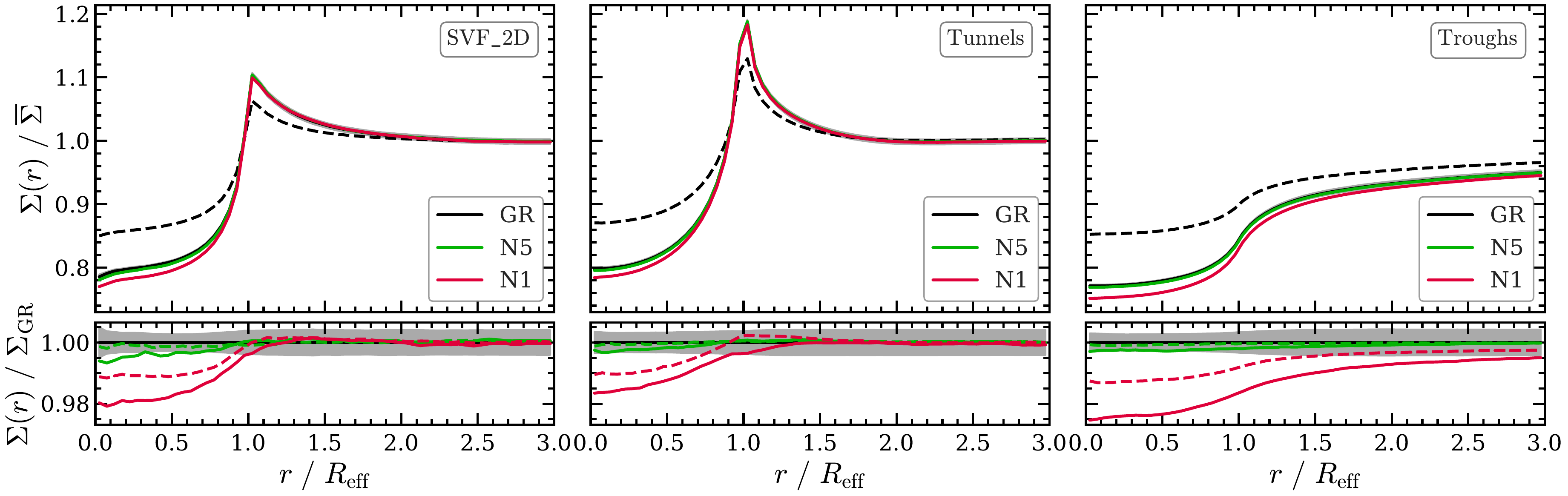}
	\end{tabular}
    \vskip -.2cm
    \caption{Void density profiles in the GR and nDGP models, for different void finding algorithms. Black, green and red lines show voids identified in the GR, N5 and N1 simulations, respectively. Solid and dashed lines correspondingly show results for $z=0.0$ and $z=0.5$. For each void finder, the lower sub-panel shows the ratio between the different gravity models and GR. The shaded regions show the bootstrap error of the mean, calculated using 5 realisations of the GR simulation.
    }
    \label{fig:void_den_profile}
\end{figure*}

\begin{figure}
	\includegraphics[width=\columnwidth]{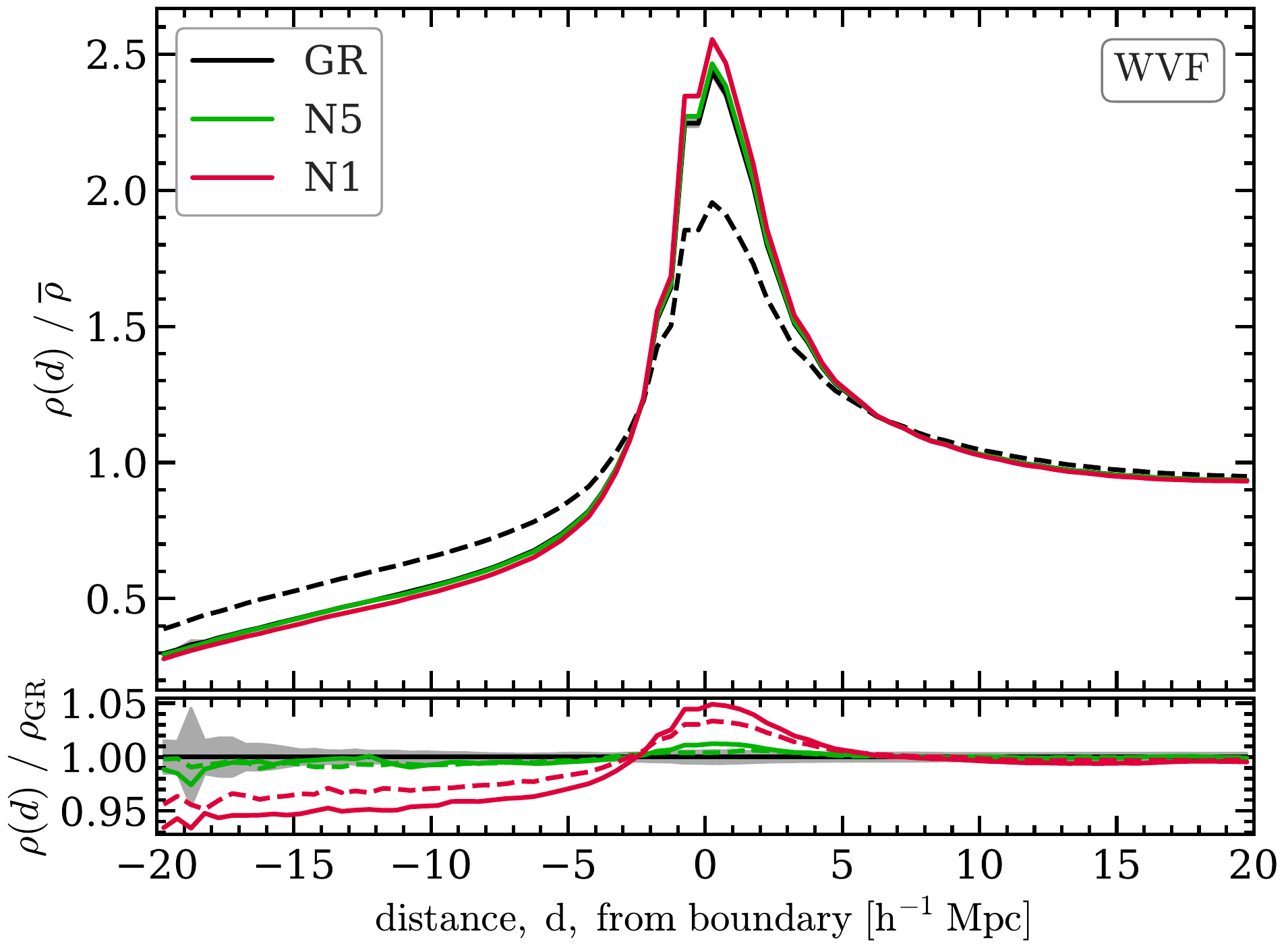}
    \vskip -.2cm
    \caption{The void density profile for the \WVF{}, but stacked with respect to the distance from the void boundary \citep{Cautun2016a}. By convention, negative and positive distances correspond to regions inside and outside the void, respectively. The void edge corresponds to a distance, $d=0$. The boundary profile naturally accounts for the void shape when stacking multiple objects and leads to a better segregation between the inner and outer regions of voids.
    }
    \label{fig:void_den_profile_boundary}
\end{figure}

\begin{figure*}
     \centering
     \begin{tabular}{cc}
        \includegraphics[width=\figwidth,angle=0]{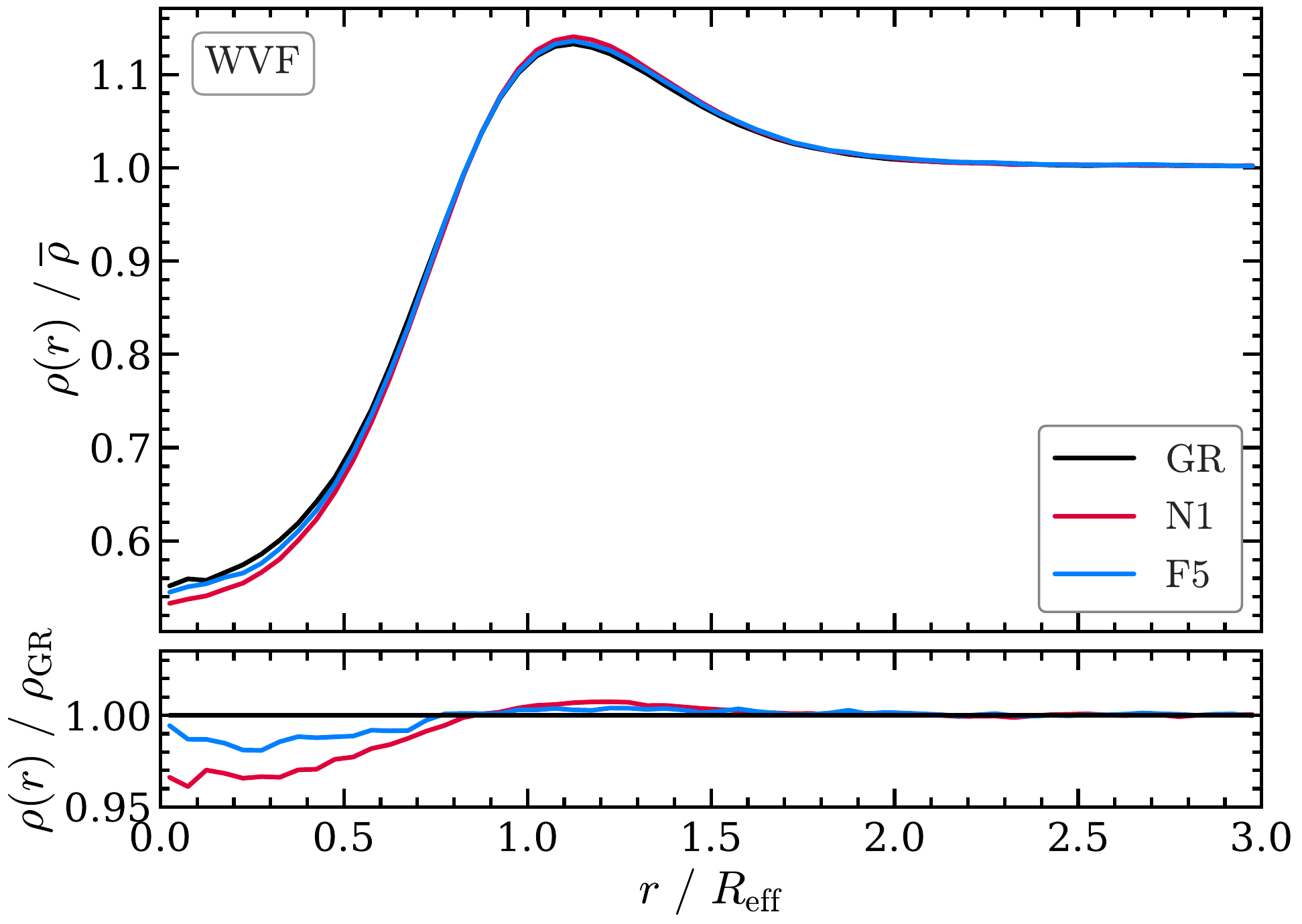} &
        \includegraphics[width=\figwidth,angle=0]{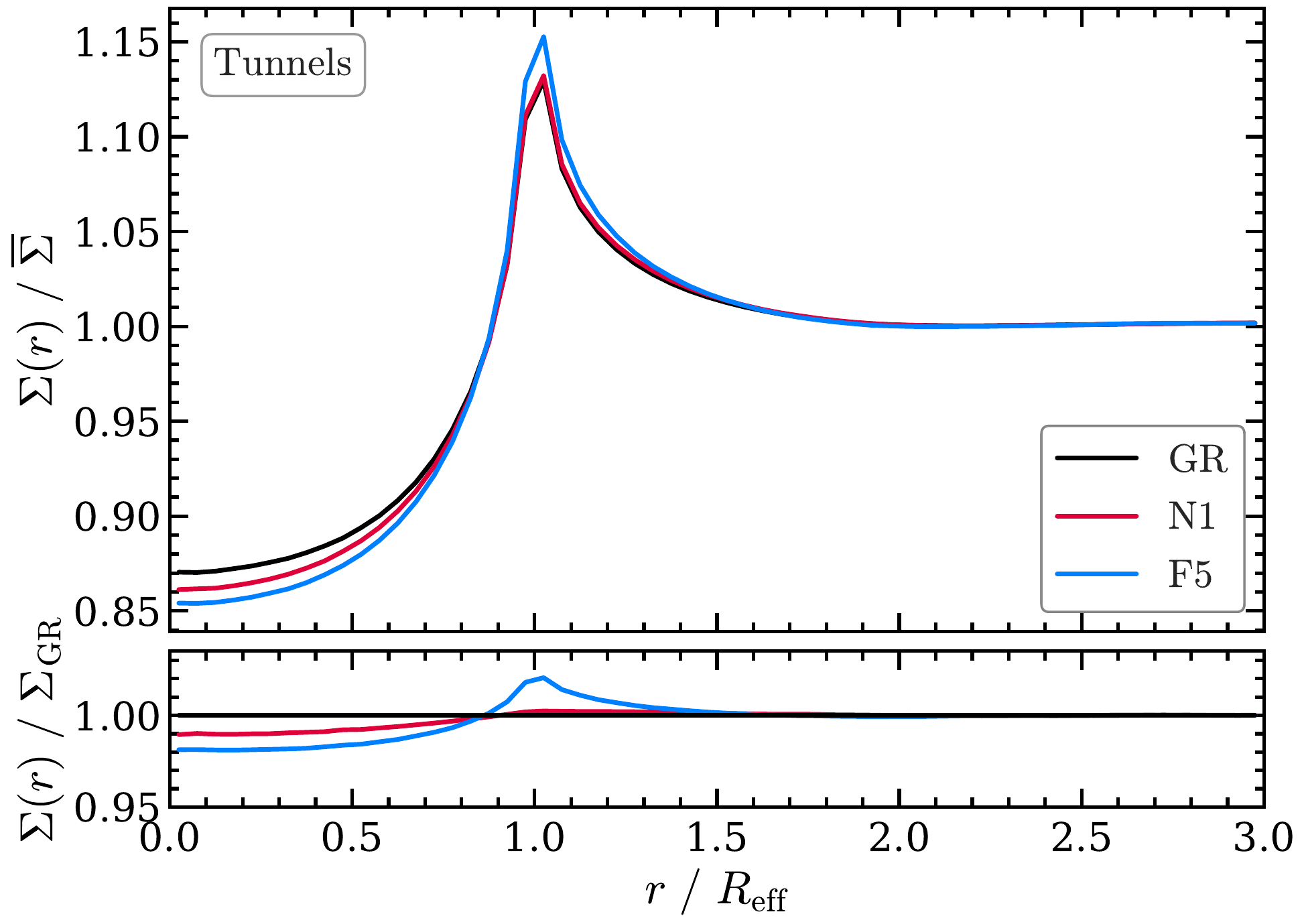}
     \end{tabular}
     \reduceVerticalOffset{}
     \caption{Void density profiles for \WVF{} voids (left) and \tunnels{} (right) in GR, nDGP and $f(R)$ gravity models (black, red and blue curves, respectively) at $z=0.5$. The upper sub-panels show the density profiles, while the lower sub-panels quantify the ratio between each model and the standard GR case.} 
     \vspace{-.2cm}
     \label{fig:void_den_profile_nDGP_fR}
\end{figure*}

\begin{figure}
	\includegraphics[width=\columnwidth]{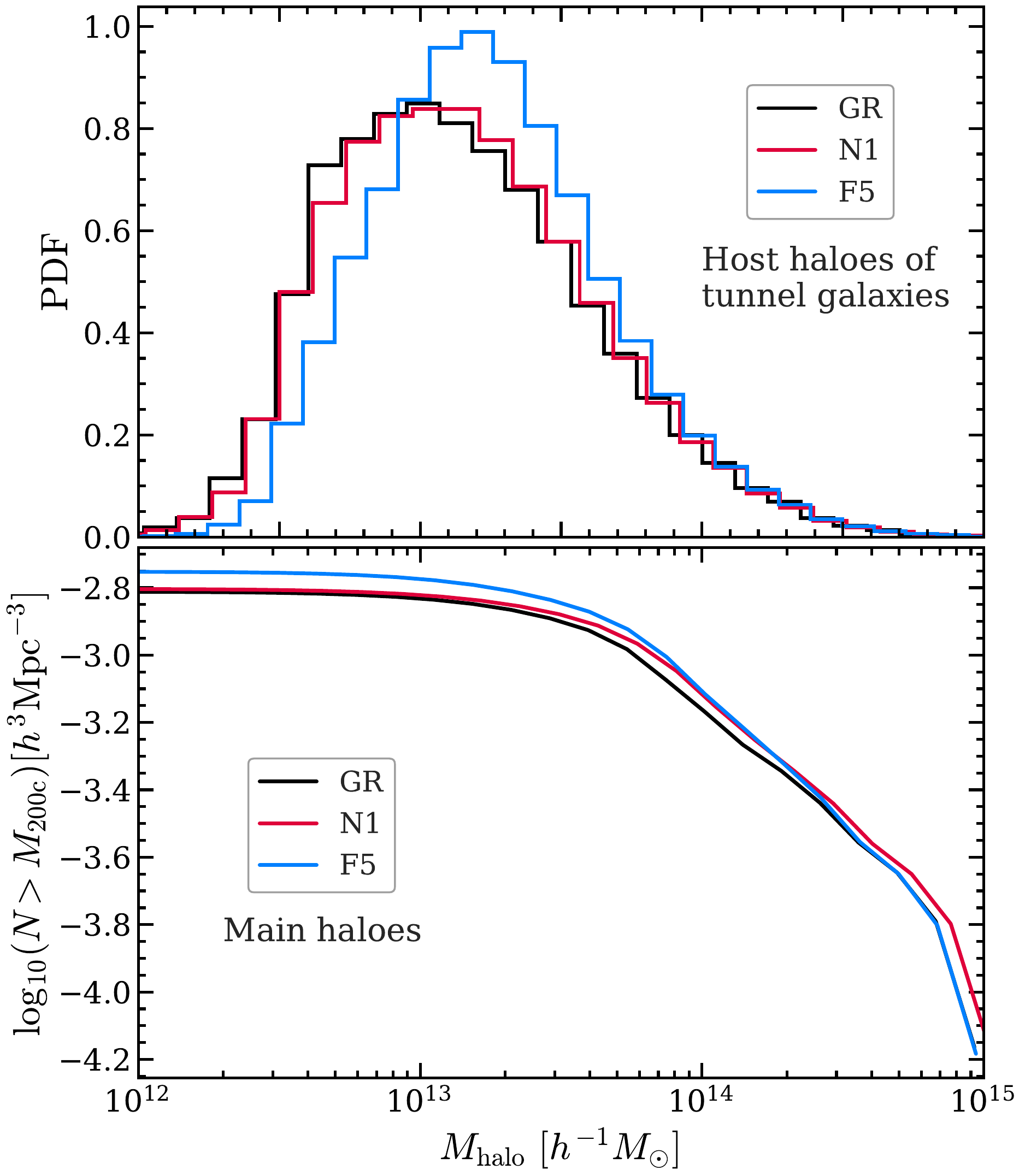}
    \vskip -.2cm
    \caption{Upper panel: the probability density function of host halo masses for HOD galaxies that participate in the definition of \tunnels{} for different gravity models at $z = 0.5$. Lower panel: the cumulative mass function for main haloes in these models.
    }
    \label{fig:HMF}
\end{figure}

An effective characterisation of cosmic voids is their density profiles. One expects to find a low matter content around void centres, and a steep increase in the matter density near the boundaries of the voids.
We measure the spherically-averaged matter density around each void in concentric radial shells of 0.05 void radii of thickness. Each profile is re-scaled by the void effective radius and normalised to the mean matter density in the simulation before stacking them.

The upper panels of Fig.~\ref{fig:void_den_profile} show the void density profiles for voids identified with the 3D finders. Different line colours indicate the different gravity models, while solid and dashed line styles show results for simulation snapshots at $z=0.0$ and $z=0.5$, respectively. For each void finder, the density profile ratio between the different models and GR is shown in the lower sub-panel. Shaded grey regions display the error around the mean.

We notice that these voids have very low densities near the centres, and a prominent ridge around $r = R_{\rm{eff}}$. At large distances from void centres, the densities converge to the mean, as expected. The over-dense ridge around the void radius is a characteristic feature of void profiles. Matter in the inner void shells evacuate faster than the outer shells, and eventually the shells cross. This results in a prominent accumulation of matter that constitutes the void walls (\citealt{Blumenthal1992,Sheth2003}). 
The \SVF{} shows a more pronounced peak compared to the \WVF{} and \zobov{}, 
which is due to spherically averaging over the non-spherical shapes of the latter.

For all the 3D finders considered here, nDGP produces more underdense voids than GR. As will be shown in the following section, the Vainshtein screening in nDGP is ineffective inside voids, giving rise to a non-zero fifth force inside them. On average, the fifth force points along the direction of the Newtonian force and thus leads to a faster evacuation of matter from voids, making them more underdense than in GR. 
This in turn results in a slightly more overdense ridge at $r\sim R_{\rm eff}$ in nDGP due to mass conservation. When comparing curves corresponding to different redshifts, we observe that the differences between nDGP and GR become larger at later times in the simulation (this is true for the overall matter distribution in general). The largest fractional differences between the models arise around the void centres, and these fluctuate around 5 per cent at $z=0.0$.

The lower panels of Fig.~\ref{fig:void_den_profile} show the projected matter density profiles for underdensities identified using 2D void finders. The \SVFtwoD{} and \tunnels{} show similar features as the 3D finders, but they are much shallower near the centre (note the different scales on the corresponding $y$-axes). These 2D voids correspond to elongated underdense structures along the line-of-sight; as we project the full simulation box into a plane for their identification, the density contrast between the most overdense and underdense parts of these voids is not as high as in the 3D case. In the case of troughs, the density ridge is not present, as they are probably embedded in underdensities that are larger than their size, having ridges that lie farther away than $3 R_{\rm{eff}}$. The \SVFtwoD{} and \tunnels{} show similar fractional differences between nDGP and GR, fluctuating around 2 per cent near the void centres. In the case of troughs, the differences between N1 and GR can be seen up to larger distances from their centres, which is due to troughs being fairly underdense even beyond the effective radius, and as such, the Vainshtein mechanism does not screen the fifth force as efficiently as in the case of \SVFtwoD{} or \tunnels{} at those distances.

As mentioned above, \WVF{} and \zobov{} voids are highly non-spherical (see e.g. Fig. \ref{fig:density_map}), which means that stacking these voids under the assumption of spherical symmetry leads to a smearing-out of the over-density ridges that delineate the voids. This explains why the density profile of these voids slowly increases towards the void edge and why they do not have a sharp density maximum at $r=R_{\rm eff}$ as in the case of \SVF{}. The non-spherical shape of voids also explains why the differences between the nDGP and the GR models depend on void identification method. By construction, \SVF{} voids have very few over-dense regions for $r<R_{\rm eff}$ and thus the difference between the void density profiles in the nDGP and GR models is nearly constant up to $r\sim R_{\rm eff}$. In contrast, for \WVF{} and \zobov{} voids, the differences between the void profiles in the nDGP and GR models are constant only for $r< 0.5 R_{\rm eff}$ and they slowly decrease for $r> 0.5 R_{\rm eff}$. Due to their non-spherical shape, many \WVF{} and \zobov{} voids contain over-dense regions inside their effective radius, $R_{\rm eff}$. These overdense regions contain more mass in nDGP models than in the GR one, and thus averaging over them reduces the difference between void profiles in nDGP and GR models. 

We can maximally preserve and use the information contained in the non-spherical shape of \WVF{} and \zobov{} voids by stacking them with respect to the distance from the void boundary, as proposed by \citet{Cautun2016a}. We illustrate the boundary stacking procedure for \WVF{} voids in Fig. \ref{fig:void_den_profile_boundary}. To discriminate between the void interior and exterior, we follow the \cite{Cautun2016a} convention and assign a negative distance to the points inside the void. The void edge corresponds to a distance of 0, while positive distances indicate points outside the void.

As can be readily noticed, the boundary profile produces a sharply peaked transition between the void interior and exterior regions. This method enhances the signal from the void ridge in the \WVF{} and highlights differences up to 5 per cent between nDGP and GR around this region, which is not observed in the smoother spherically-averaged density profile (middle upper panel of Fig. \ref{fig:void_den_profile}). Furthermore, inside the void we find a nearly constant difference between the void density profiles in the nDGP and GR models, with the former having emptier voids. A similar behaviour is expected when applying this procedure to \zobov{} voids, although the explicit calculation is not done in this work.

The \WVF{} boundary profiles of Fig.~\ref{fig:void_den_profile_boundary} illustrate very nicely the phenomenology of nDGP models. The fifth force enhances the action of gravity in voids and leads to a faster transport of mass out of voids. This can also be seen in the radial velocity profiles of voids, with nDGP voids having stronger outflows than GR ones \citep{Falck2017}, which is qualitatively similar to the effect of the fifth force in $f(R)$ gravity \citep{Cai2014}. Due to mass conservation, the displaced mass is deposited at the void edges, leading to more massive filaments and nodes.

Fig.~\ref{fig:void_den_profile_nDGP_fR} compares the void density profiles in nDGP with those in $f(R)$ gravity models calculated in \citet{Cautun2017}. The left- and right-hand side panels show results for the \WVF{} and \tunnels{}, respectively. Similarly to nDGP, $f(R)$ predicts more underdense voids than GR. The nDGP N1 model and the $f(R)$ F5 model have similar $\sigma_8$ values at $z=0$, however they predict different void profiles. Compared to N1, voids in F5 are less underdense in their interiors for the \WVF{}, while the opposite is found for \tunnels{}. Furthermore, the void ridge of \tunnels{} is more prominent in $f(R)$ gravity than in nDGP.

We find that nDGP is more effective at evacuating 3D voids than $f(R)$ gravity. This is possibly a consequence of the fact that the fifth force in the former is long-ranged, which means that matter deep inside large voids still feels the enhanced attractive force from matter outside the voids. On the other hand, the fifth force in $f(R)$ gravity is short-ranged and decays exponentially beyond the Compton wavelength of the scalar field\footnote{In the Jordan frame, the trace of the modified Einstein equation in $f(R)$ gravity gives a second-order different equation for $f_R={\rm d}f(R)/{\rm d}R$,
\begin{equation}
    \Box f_R = \frac{1}{3}\left[R(f_R)-f_RR(f_R)+2f(f_R)+8\pi G\rho_m\right],
\end{equation}
where $f_R$ is considered as a scalar dynamical degree of freedom (the {\it scalaron}), and $R$, $f$ are expressed as functions of $f_R$. This places an implicit requirement that the function $f_R(R)$ can be inverted to write $R(f_R)$, which is satisfied by the \citet{Hu2007} $f(R)$ model studied here.}, which can be smaller than the size of the void itself. As a result, the fifth force from matter outside the voids may not be able to propagate deep inside voids. We shall show this contrast between nDGP and $f(R)$ gravity more explicitly in the next subsection when describing the void force profiles. 

We observe the opposite trend for \tunnels{}, which are emptier in F5 than in N1, cf.~the right panel of Fig.~\ref{fig:void_den_profile_nDGP_fR}. In order to understand this better, let us consider the distribution of host halo masses for the HOD galaxies that define the \tunnels{}' boundaries (what we henceforth refer to as `tunnel galaxies'), shown in the upper panel of Fig. \ref{fig:HMF}. We observe that, on average, the host halo mass of tunnel galaxies in F5 is larger than in N1 and GR. This is also true for the host halo mass distribution of the full galaxy sample, indicating that in principle, galaxies defining tunnels are rather typical.

Here we comment on a key difference between 3D (such as \WVF{}) voids and 2D voids such as tunnels: for the former, their boundaries usually coincide with sheets and filaments, and their density ridges can be naturally explained by the accumulation of the evacuated matter from void interiors near their edges; for the latter, there is not a similar coincidence between their boundaries and filaments (indeed it is likely that sheets and filaments with unresolved haloes cross their interiors, cf.~Fig.~\ref{fig:tunnels_sketch}) -- therefore we expect that their density ridges receive major contributions from the host haloes of tunnel galaxies (and possibly small unresolved structures around these host haloes). This offers an explanation to the more prominent ridges for F5 \tunnels{} in the right-hand side of Fig.~\ref{fig:void_den_profile_nDGP_fR}, namely it is a simple consequence of the fact that in F5 the host haloes of tunnel galaxies are more massive than in GR and N1. Due to mass conservation, the interior of these \tunnels{} will also be more underdense when compared to N1 and GR. 

The differences between the distribution of host halo masses for tunnel galaxies can be explained more clearly by looking at the lower panel of Fig.~\ref{fig:HMF}, which shows the cumulative halo mass function (HMF) for each gravity model. We can see that massive haloes in nDGP are more abundant than in F5, and the N1 HMF is higher than the F5 one all the way down to $\sim 10^{14}~h^{-1}M_{\rm{\odot}}$, while the situation reverts for lower masses. This is a consequence of how the growth of haloes is affected by the different screening mechanisms operating in these models. On the one hand, many small haloes are in relatively low density regions with only a small reservoir of matter around them. They grow more slowly, by accreting matter from their nearby surroundings. In chameleon models, the fifth force is unscreened in low-density regions, which can boost the accretion rate; in Vainshtein models, these small haloes can self-screen (their Vainshtein radii are generally significantly larger than their size); they do not grow as fast as in chameleon models. On the other hand, large haloes (i.e. those more massive than ${\sim}10^{14}~h^{-1}M_\odot$) form by concentrating the matter within regions of radius of $\sim 10~h^{-1}\rm{Mpc}$ \citep[see, e.g., Fig.~1 of][]{Zentner:2006vw}: for chameleon models this is larger than the Compton wavelength of the scalar field, so the fifth force (which is also often well screened in high-density regions) is ineffective at attracting more matter from even larger regions; in Vainshtein models, the fifth force is long-ranged, which helps migrate more matter from beyond $\sim10~h^{-1}\rm{Mpc}$ into the central halo-formation region
-- this offers a larger effective reservoir of matter for the halo to grow from. Looking back at the upper panel of Fig.~\ref{fig:HMF}, we see that most of the tunnel galaxies have host halo masses around $\sim 10^{13}~h^{-1}M_{\rm{\odot}}$ -- this is precisely the range of halo masses where the HMF in N1 is more similar to GR than F5.

The large difference in the dark matter density profiles around tunnels in N1 and F5 (Fig.~\ref{fig:void_den_profile_nDGP_fR}) -- which is, as we shall see later, the reason for strong differences in the tunnel lensing tangential shear profiles in these models -- can therefore be understood as originating from the different shapes of the HMFs near $10^{13}~h^{-1}M_\odot$, which corresponds to the mass of the typical halo which hosts a galaxy for a galaxy number density, $n_g\sim3\times10^{-4}~(h^{-1}{\rm Mpc})^{-3}$.

If this explanation is correct, then the following comments may apply, which will leave interesting possibilities to be explored in future works: 
\begin{enumerate}
   \item The fact that HOD galaxies are hosted by haloes of different masses in the various models studied here is a consequence of the differing HMF shapes in these models {\it and} the requirement that the different models have the same galaxy clustering (which in turn is the result of constraints from observations). This leaves little leeway to modify the HOD, making the results of tunnel density and lensing profiles robust.
   \item In this sense, the use of tunnels is effectively a way to `select' dark matter haloes from the complete halo catalogue. For tunnels we find that the selected haloes are representative of the whole catalogue of haloes hosting HOD galaxies, 
   though other ways to select haloes (e.g., based on the environmental galaxy number density) are possible. One still needs to find suitable statistics of the halo masses, with tunnel lensing to be shown below being one example. Other examples include the stacked galaxy-galaxy lensing signal around tunnel galaxies (which directly probes the halo lensing masses), or redshift space distortions around the host haloes of tunnel galaxies (this has the theoretical advantage that these haloes are likely to be unscreened in F5 so that they will show a stronger difference of dynamical masses from GR or N1). Note that this way of `selecting' haloes also does not require redshift information because tunnels are defined by the projected galaxy density field. 
   \item If for a different $f(R)$ model the shape of its HMF differs from GR at some smaller halo mass scale, one can choose a galaxy catalogue with higher number density (which means that the most probable host halo mass shifts to smaller values as well) to find the sweet spot for testing that theory. However, a caveat is that for smaller haloes the lensing signal may be weaker as well, so a more detailed study is needed to see if this can offer a strong test for weaker $f(R)$ models, such as F6 or F5.5 (|$f_{R0}|=10^{-5.5}$).
\end{enumerate}

\subsection{Force profiles} \label{subsec:void_force_profiles}

\begin{figure*}
     \centering
     \begin{tabular}{cc}
        \includegraphics[width=\figwidth,angle=0]{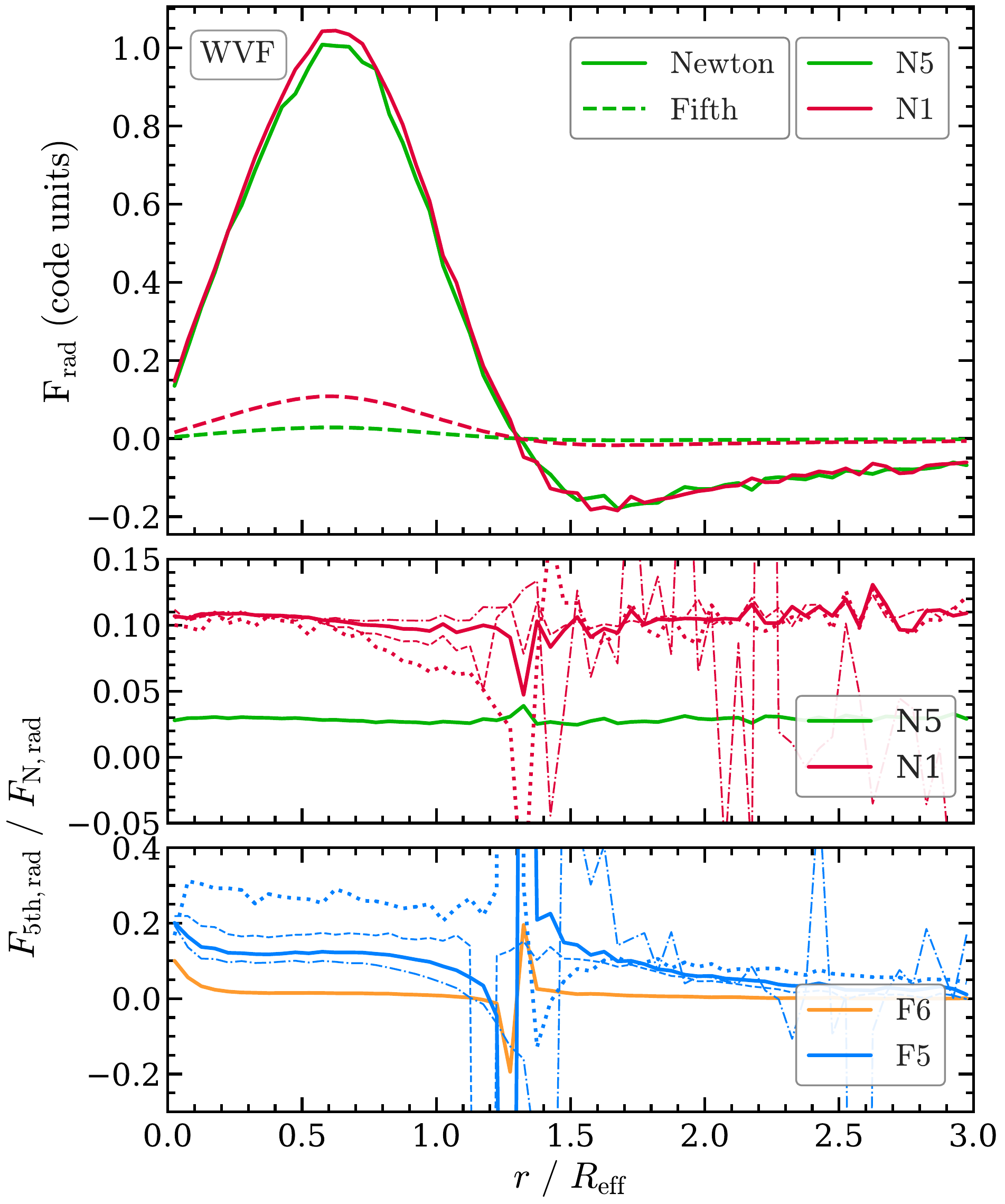} &
        \includegraphics[width=\figwidth,angle=0]{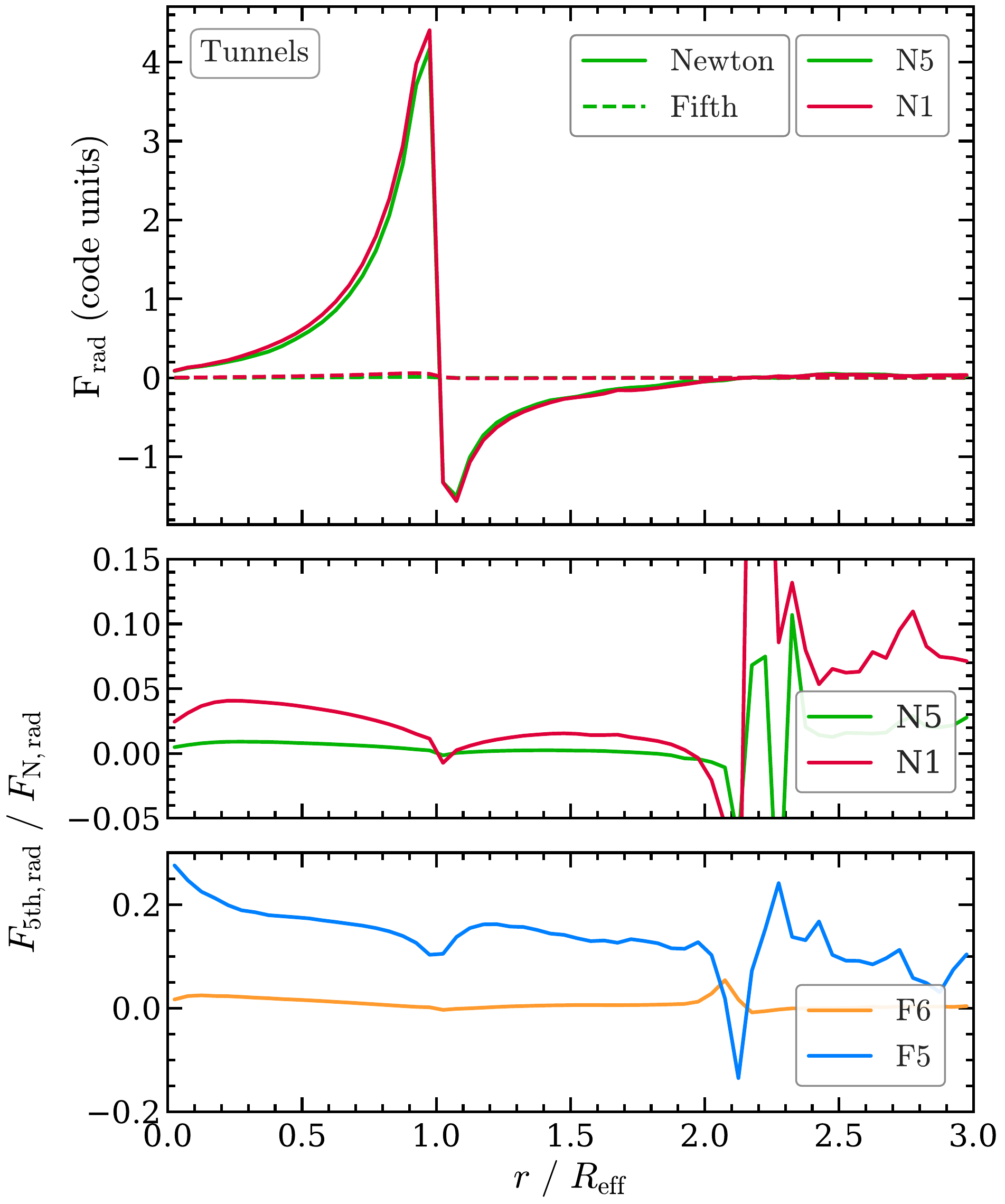}
     \end{tabular}
     \reduceVerticalOffset{}
     \caption{Force profiles for WVF{} voids (left) and \tunnels{} (right). Top row: The radial component of the force, with respect to the void centre. The Newtonian and fifth forces are shown by the solid and dashed lines, respectively. Middle row: The ratio between the radial Newtonian and fifth forces in nDGP. Bottom row: Same as the second row, but for the $f(R)$ gravity models.
     The bottom two panels in the left-hand column show the average force profiles for all the voids (solid line) as well as the force profiles for \WVF{} voids of different sizes, with $R_{\rm eff}\in [0,20]$ (dotted line), $[20,35]$ (dashed line) and $[35,100]~h^{-1}$Mpc (dashed-dotted line). For clarity, the force profiles for voids of different sizes are shown only for the N1 and F5 models.
     }
     \vspace{-.2cm}
     \label{fig:void_force_profile}
\end{figure*}

As was seen in the previous section, matter is distributed differently around nDGP and $\Lambda$CDM voids. In this section we 
explore in greater detail the physical mechanism that drives these differences in the void density profiles.

We obtain values of the Newtonian and fifth forces felt by particles directly from the simulation output. To do so, we calculate the average radial component, with respect to the void centre, of the forces felt by particles in each shell (in the case of 2D voids, we also project the force vector onto the plane-of-the-sky). We then proceed to stack the profiles in the same way as we did for the density profiles (re-scaling them by the void radius and weighing each profile by $R_{\rm{eff}}^2$).

Fig.~\ref{fig:void_force_profile} shows the void force profiles for the \WVF{} and \tunnels{} at $z=0.5$. The radial component of the Newtonian and fifth forces is shown in the top row by the solid and dashed lines, respectively. Positive values indicate forces that point away from the void centre. For the \WVF{} the Newtonian force is always positive for $r < R_{\rm{eff}}$, meaning that matter is being evacuated from the voids. The fifth force behaves in a similar way, but with a smaller magnitude. The presence of the fifth force inside the voids in nDGP contributes to push matter outwards and evacuate it faster than in GR. This confirms our explanation for the differences seen in the 3D void density profiles of the previous section, that nDGP voids are more underdense in their interiors and have larger overdensities at their ridges. For $r > R_{\rm{eff}}$, the \WVF{} force profiles turn negative, indicating that matter is being pulled towards the voids at such distances. This is due to the overdense void walls that attract matter towards them. In the case of \tunnels{}, the force profiles show a similar behaviour, but there is a much sharper transition from positive to negative values at the void radius, in agreement with their sharper density contrast seen in the density profiles of Fig.~\ref{fig:void_den_profile}.

The second row in Fig. \ref{fig:void_force_profile} shows the ratio between the radial components of the fifth and Newtonian forces in nDGP. For \WVF{} voids, the ratios are relatively constant at around 10 and 3 per cent for N1 and N5, respectively, while for \tunnels{} the radial force ratios are much smaller, dropping below $4$ per cent for N1. This can be understood as follows: in the case of \tunnels{}, we are projecting along the line-of-sight and, for most part, the forces acting on the particles at very different distances to the observer have random directions, so that they largely cancel out when taking the radial average. The particles at similar line-of-sight distances as the haloes at the tunnel boundaries, on the other hand, have coherent directions of their forces. The latter particles dominate the stacked value of the radial force. These particles are more likely to be within the Vainshtein radius of the massive haloes that are used to define tunnels and therefore be screened or partially screened.

The third row in Fig.~\ref{fig:void_force_profile} shows the ratio between the radial force components for two $f(R)$ models, F5 and F6 \citep{Cautun2017}, for comparison. The \WVF{} force ratios in F6 and F5 are larger than those found in N5 and N1, respectively. For both $f(R)$ models, the force ratios show a vertical asymptote at $r\approx 1.3 R_{\rm{eff}}$ which corresponds to the fact that radial fifth force switches from positive to negative at slightly lower $r$ values than the Newtonian force. For \tunnels{}, the force ratios have similar values to the 3D case, in contrast to what is observed for nDGP. The transition from being mostly screened to mostly unscreened is very sharp in $f(R)$ gravity while much more gradual in nDGP (e.g. compare Figs. 7 and 8 in \citealt{Falck2015}). This means that a screened halo in $f(R)$ gravity is likely to be surrounded by unscreened regions, while a screened halo in nDGP can usually at least partially screen its surrounding environment up to ${\sim}10$ times the halo radius. Thus, while the haloes at the tunnels edges in $f(R)$ might be screened (see the small dip in force ratio at $r=R_{\rm{eff}}$), the regions around these haloes (including the interiors of the tunnels) are mostly unscreened.

One observation from the lower left panel of Fig.~\ref{fig:void_force_profile} is that the fifth force ratio is significantly below $1/3$ in voids. This seems to contradict the naive expectation that in $f(R)$ gravity the fifth force has $1/3$ the strength of Newtonian gravity in low-density regions such as voids. To understand this, we have also recomputed the relations in  Fig.~\ref{fig:void_force_profile} by splitting our \WVF{} void catalogues into three bins with different effective void radii: $R_{\rm eff}\in [0,20]$, $[20,35]$ and $[35,100]~h^{-1}$Mpc. These bins are shown by the dotted, dashed and dot-dashed lines in the second and third rows of the left panels of Fig.~\ref{fig:void_force_profile}. We find that, as expected, the fifth force ratio in nDGP is almost independent of the void size, while for F5 the force ratio has an average value of $\sim0.33$, $\sim0.2$ and $\sim0.1$ respectively for the three bins. We interpret this result as a consequence of the short-range nature of the fifth force in chameleon models: for the smallest void radius bin, the void radius is smaller than the scalar field Compton wavelength, so that the fifth force from the surrounding matter of the voids can reach the void centres un-decayed; while for the larger void radius bins the fifth force by matter outside voids decays and cannot be felt by matter deep inside voids. This can also explain why F5 is less effective in evacuating 3D voids than N1.

\subsection{Lensing tangential shear profiles} \label{subsec:void_lensing_profiles}

\begin{figure*}
    \begin{tabular}{c}
	    \includegraphics[width=\textwidth]{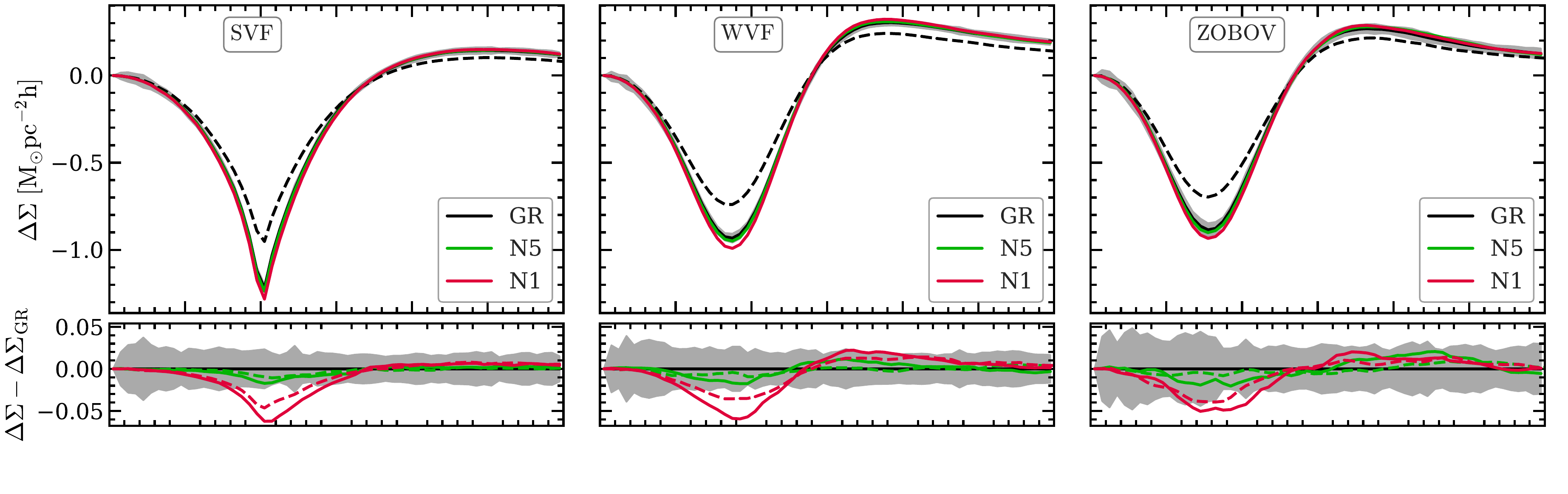} \verticaloffset
	    \includegraphics[width=\textwidth]{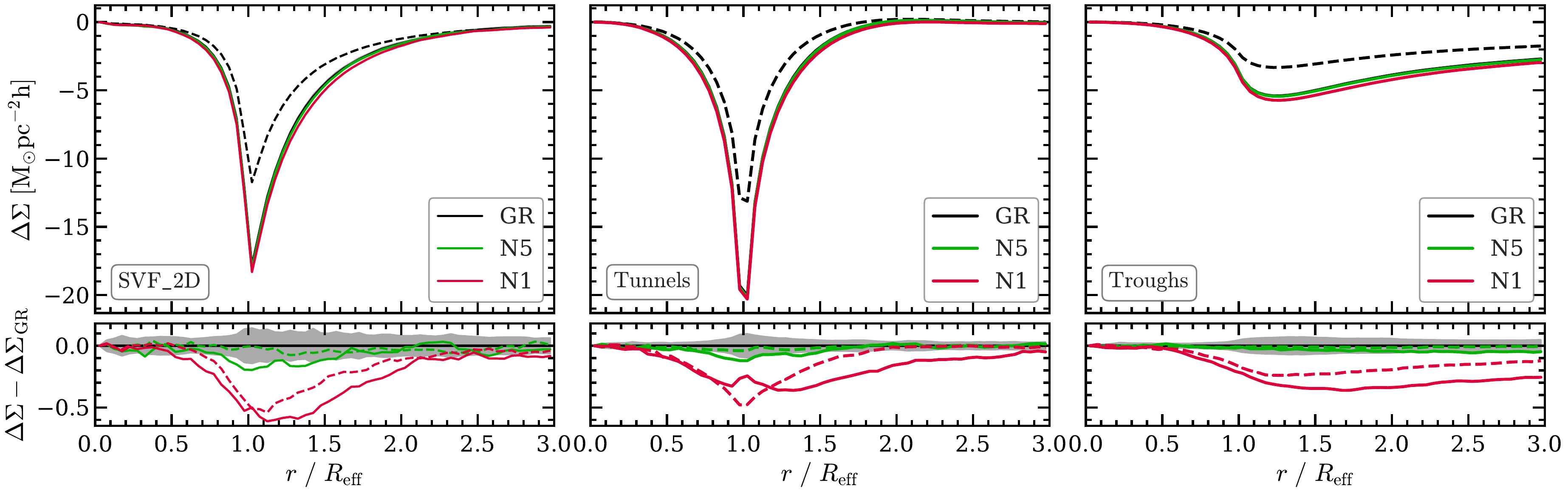}
	\end{tabular}
    \vskip -.2cm
    \caption{The upper sub-panels show the void surface mass density profiles in GR and nDGP models, while the lower sub-panels show the absolute difference between each model and GR. Colours and symbols are the same as in Fig. \ref{fig:void_den_profile}. The shaded grey regions show the uncertainties associated to the GR signal. For the 2D voids, the errors correspond to the standard deviation for a $(1.024 h^{-1}\rm{Gpc})^3$ volume, which is the volume of each of our simulation boxes. For 3D voids, we re-scale the errors to show the standard deviation for a \Euclid{}-like survey with a $10(h^{-1}\rm{Gpc})^3$ volume, which is 9.3 times larger than the volume of our simulations.
    }
    \label{fig:void_lensing_profile}
\end{figure*}

\begin{figure}
	\includegraphics[width=\columnwidth]{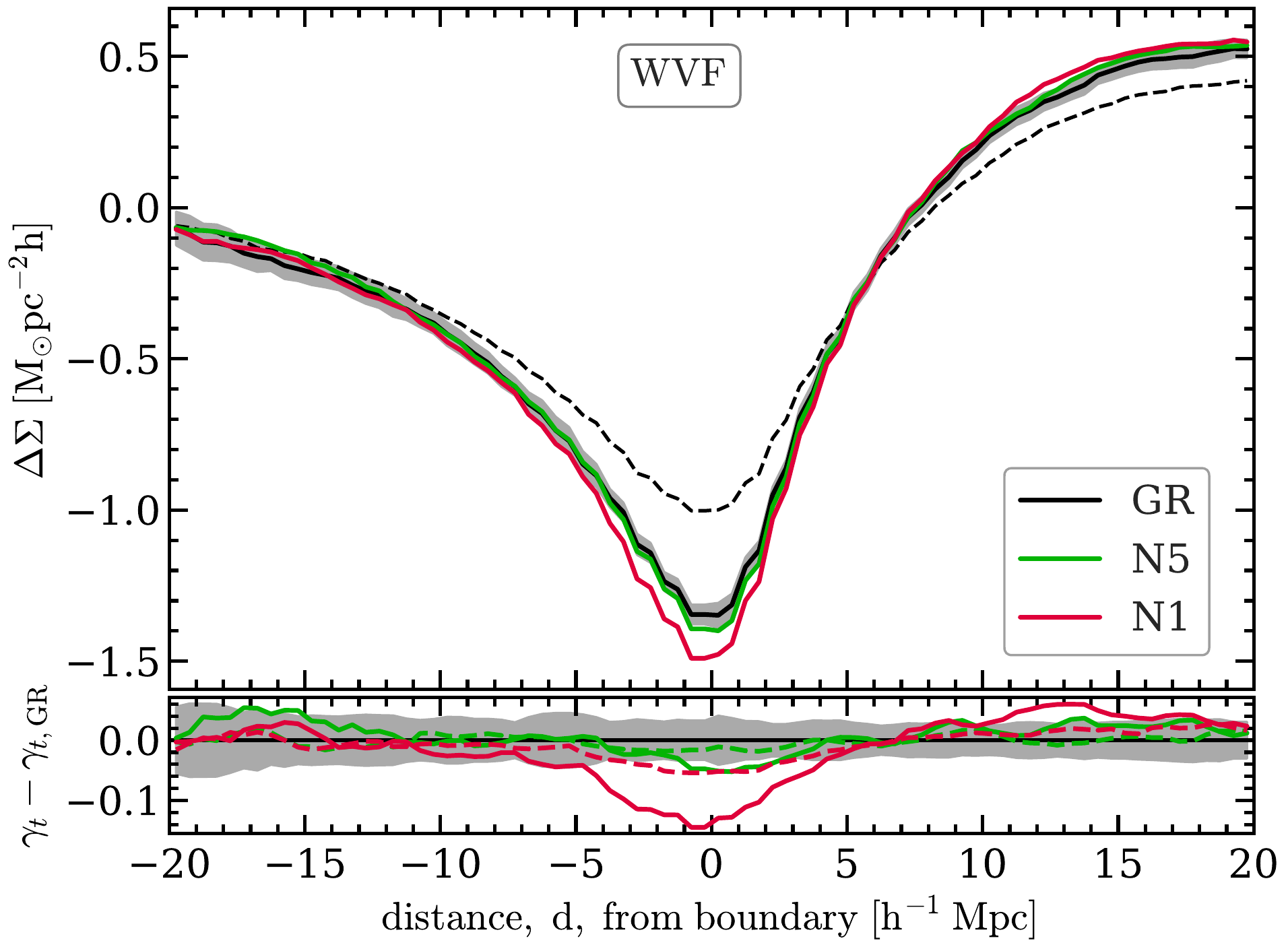}
    \vskip -.2cm
    \caption{The differential surface mass density profile of \WVF{} voids stacked according to the distance from the void boundary. See Fig. \ref{fig:void_den_profile_boundary} for more details.
    }
    \label{fig:void_lensing_profile_boundary}
\end{figure}

\begin{figure*}
	\includegraphics[width=\textwidth]{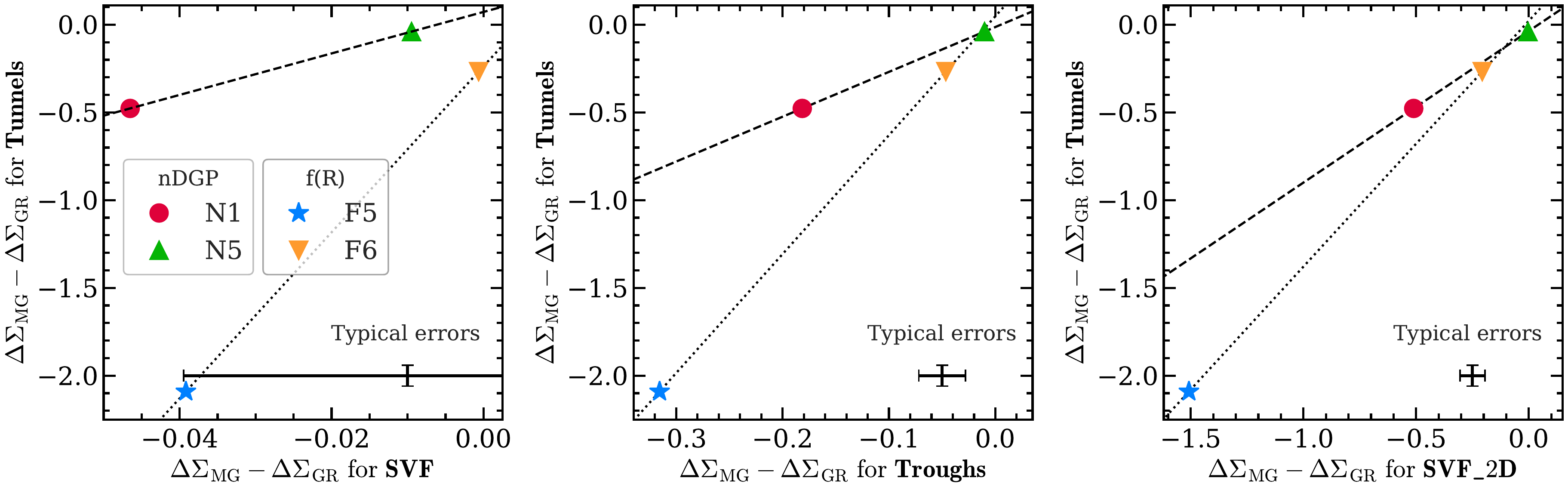}
    \vskip -.2cm
    \caption{Comparison of the void lensing signal of two modified gravity theories, nDGP and $f(R)$ gravity. Each axis shows the difference in surface mass density, $\Delta\Sigma_{\rm MG}-\Delta\Sigma_{\rm GR}$, between a modified gravity model and GR measured at the void radius, $r=R_{\rm{eff}}$ (see bottom panels in Fig.~\ref{fig:void_lensing_profile}).
    The vertical axes always show results for \tunnels{}, while the horizontal axes show results for the \SVF{}, \troughs{} and \SVFtwoD{} respectively from left to right. Different symbols correspond to different gravity models, with the dashed and dotted lines connecting specific models in the same class. The error bars in each panel show the typical $\Delta\Sigma_{\rm MG}-\Delta\Sigma_{\rm GR}$ errors predicted for an {\sc lsst}-like survey.
    The figure shows that nDGP and $f(R)$ models occupy distinct regions of the plot and that, were we to measure a deviation from GR, combining lensing measurements from two different voids allow us to distinguish these two different modified gravity theories.
    }
    \label{fig:lensing_nDGP_vs_fR}
\end{figure*}

Gravitational lensing is the deflection of light and distortion of the image of a background source by a foreground lens object. In the strong lensing regime, very concentrated mass distributions can bend the light from background galaxies in such a way as to produce giant arcs and multiple images. These configurations are rare, and most lensing events in the Universe instead occur in the weak lensing regime, where the light deflection from a single source is hardly noticeable. In spite of this, the collective analysis of many such events can produce a measurable signal. Indeed, the sharp density contrast near the boundaries of cosmic voids produces a detectable weak lensing signal when many voids are stacked together \citep[e.g.][]{Clampitt2015,Sanchez2017,Davies2018}. In our case, the quantity to be measured is the shear of the shape of background galaxies by the mass distribution around voids. This can be quantified by the void tangential shear, $\gamma_{t}(r)$, which is proportional to the excess of projected mass density along the line of sight:
\begin{equation}
\gamma_t(r) \Sigma_c = \Delta \Sigma(r) = \overline{\Sigma} (<r) - \Sigma(r)\ .
\end{equation}
Here $\overline{\Sigma} (<r)$ is the mean enclosed surface density within $r$, and $\Sigma_c$ is the critical projected mass density, defined as:
\begin{equation}
\Sigma_c = \frac{c^2}{4 \pi G} \frac{D_s}{D_l D_{ls}}\ ,
\end{equation}
where $c$ is the speed of light and $G$ is the gravitational constant. $D_l$, $D_s$ and $D_{ls}$ are the angular diameter distances between lens-observer, source-observer and source-lens, respectively. 

For 2D voids, $\Sigma(r)$ is simply given by the profiles shown in the lower panels of Fig.~\ref{fig:void_den_profile}. To compute $\Sigma(r)$ for 3D voids, we integrate their density profile  $\rho(r)$ (upper panels of Fig. \ref{fig:void_den_profile}) using an integral along the line-of-sight (\citealt{Barreira2015,Cai2016}):
\begin{equation} \label{eq:Sigma_r_integrated}
\Sigma(r) = \int_{-L}^{L} \rho(\sqrt{r^2 + l^2})\, {\rm d}l\ .
\end{equation}
The line-of-sight length, $L$, has to be large enough to account for correlations in the large-scale distribution of matter. Here we take $L$ to be 3 times the void radius, as the matter density profiles are converged to the mean at such distances. While this calculation is inexpensive and useful to characterise theoretical differences between different models, the dispersion around the mean of these profiles under-estimates the true uncertainty associated to $\Sigma(r)$; we know that the projected distribution of matter around each void varies significantly depending on the chosen viewing direction, but $\rho(r)$ is an average over all possible lines-of-sight, and thus does not take into account this source of error that would be present for a real observer at a fixed location.

Considering the discussion above, we calculate the mean $\Sigma(r)$ for GR and nDGP using Eq.~(\ref{eq:Sigma_r_integrated}), but the sample variance is calculated using the projected distribution of matter around each void. The covariance matrix is estimated using 500 bootstrap re-samples from 5 GR realisations, each further divided into 64 regions, similar to the calculation performed for the matter density profiles.

The differential surface mass density profiles for 3D voids are shown in the top row of Fig.~\ref{fig:void_lensing_profile}. The profiles show negative values at $r < 2 R_{\rm eff}$, highlighting the under-dense nature of these voids, so as to produce divergent lensing. As discussed in Sec.~\ref{subsec:void_den_profiles}, the \SVF{} voids have a strong density contrast at $r=R_{\rm eff}$, which results in a higher and sharper lensing signal than \WVF{} and \zobov{} voids around this region. The lower sub-panels show the absolute differences between the lensing signals of nDGP and GR models. While N5 is only mildly distinguishable from GR when compared to the size of the error bars, N1 shows a more significant difference, which is similar in magnitude for all the 3D void finders considered here.

The bottom row of Fig.~\ref{fig:void_lensing_profile} shows the differential surface mass density profiles for 2D voids. The \SVFtwoD{} and \tunnels{} show profiles that are sharply peaked at the void radius, while the \troughs{} have a smoother profile, which attains maximum value at $r\simeq 1.2R_{\rm eff}$. Compared to the 3D case, the 2D voids exhibit a much stronger lensing signal, as these voids were identified as underdese lines-of-sight in the galaxy distribution and are more likely to also correspond to substantial line-of-sight mass under-densities. The 2D voids also show stronger differences between nDGP and GR, both in absolute value and when compared to the size of the associated uncertainties. 

The differences between nDGP and GR grow with time, being the largest at the present time, which is a consequence of the longer time period in which the fifth force was active. The void lensing difference between N1 and GR is only slightly larger at $z=0.0$ than at $z=0.5$, but the difference between N5 and GR is a factor of a few larger at $z=0.0$ compared to $z=0.5$. However, the statistical error will increase by using low-$z$ voids due to the decrease of survey volume. Further investigation is needed to find out the optimal redshift window to maximise the signal to noise for distinguishing the model from GR.]

Fig.~\ref{fig:void_lensing_profile_boundary} shows the tangential shear profile of \WVF{} voids stacked according to the distance from the void boundary (similar to Fig.~\ref{fig:void_den_profile_boundary}, but now applied to $\Delta\Sigma(r)$). To calculate this signal, we follow the \citet{Cautun2016a} procedure and slice each void through its centre with a plane that is perpendicular to the line of sight. The intersection between this plane and the 3D void boundary gives a 2D closed curve which represents the void boundary in the mock plane of the sky, which we then use to calculate distances to the cells of a fine grid that covers the mock plane of the sky. We then estimate the lensing potential in the plane of the sky using the thin lens and Born approximations (see \citealt{Cautun2016a} for more details), and calculate the tangential shear on the fine grid.

By stacking \WVF{} voids with respect to their boundary, which accounts for their non-spherical shape, we find a tangential shear signal that is boosted by a factor of $\sim 1.5$ compared to the standard spherical stacking approach (compare Fig. \ref{fig:void_lensing_profile_boundary} to the top-row middle panel of Fig. \ref{fig:void_lensing_profile}). The void lensing differences between nDGP and GR are also increased, reaching differences that are even larger than those found for \SVF{} voids. 
This result, together with those presented in Fig. \ref{fig:void_den_profile_boundary}, show the increased information contained in void profiles when accounting for their non-spherical shape. However, the boundary stacking of 3D voids does not increase their weak lensing signal enough to compete with the power of 2D voids to discriminate between nDGP and GR models. The latter still show the largest difference in tangential shear profiles between nDGP and GR models, both in terms of absolute difference and when normalised by the size of their associated uncertainties.

Compared with GR, both nDGP and $f(R)$ gravity predict emptier voids, and thus a stronger void lensing signal. Both models have free parameters that control the size of the deviation from GR (e.g., compare the nDGP N1 and N5 models in Fig.~\ref{fig:void_lensing_profile}). One question is, if we were to measure a deviation from the GR prediction of void lensing, whether we can tell to which modified gravity theory it corresponds. This would be difficult to do using the lensing signal of a single void finder, since in many cases the same deviation from GR corresponds to one set of nDGP parameters and to another set of $f(R)$ parameters. However, given the qualitative difference in the behaviours of the nDGP and $f(R)$ models (see Figs. \ref{fig:void_den_profile_nDGP_fR} and \ref{fig:void_force_profile}), it may be possible to disentangle the two theories by comparing the lensing signal of different void finders. This is demonstrated in Fig.~\ref{fig:lensing_nDGP_vs_fR} which compares the lensing signal of various void finders in nDGP and $f(R)$ models. The three panels of the figure, from left to right, correspond to: tunnels versus \SVF{}, tunnels versus troughs, and tunnels versus \SVFtwoD{}. For each void finder, we show the amplitude of the lensing signal deviation from GR measured at $r=R_{\rm eff}$ for the two nDGP models studied here, i.e., N1 and N5, and for the two $f(R)$ models studied in \citet{Cautun2017}, i.e. F5 and F6. To help guide the eyes, we connect by a dashed line the two nDGP models and by a dotted line the two $f(R)$ models.

The left panel in Fig.~\ref{fig:lensing_nDGP_vs_fR}, which compares lensing by \tunnels{} and \SVF{}, shows that nDGP and $f(R)$ models seem to populate different regions of the plot, suggesting that the combination of these two void finders might be helpful to disentangle the two gravity models. Nevertheless, the error associated to the measurement of the weak lensing signal for the \SVF{}, shown in the bottom right corner, is prohibitively large for a reliable test of such theories (this is the error that would be expected for an LSST-like survey; see Sec.~\ref{sec:predictions_surveys} for more details). By looking at the middle and right-hand side panels, it appears that the \tunnels{} vs \troughs{} scenario is particularly promising, as not only nDGP and $f(R)$ gravity populate distinct regions of the plot, but also the uncertainty associated to the weak lensing signal from the \troughs{} is much smaller than the difference in surface mass density between the gravity models. The \tunnels{} versus \SVFtwoD{} are in a similar situation, although the surface mass density values for N5, F6 and F5 are closer together in parameter space when compared to the other void finder combinations.

\section{Predictions for EUCLID and LSST} \label{sec:predictions_surveys}

\begin{figure}
	\includegraphics[width=\columnwidth]{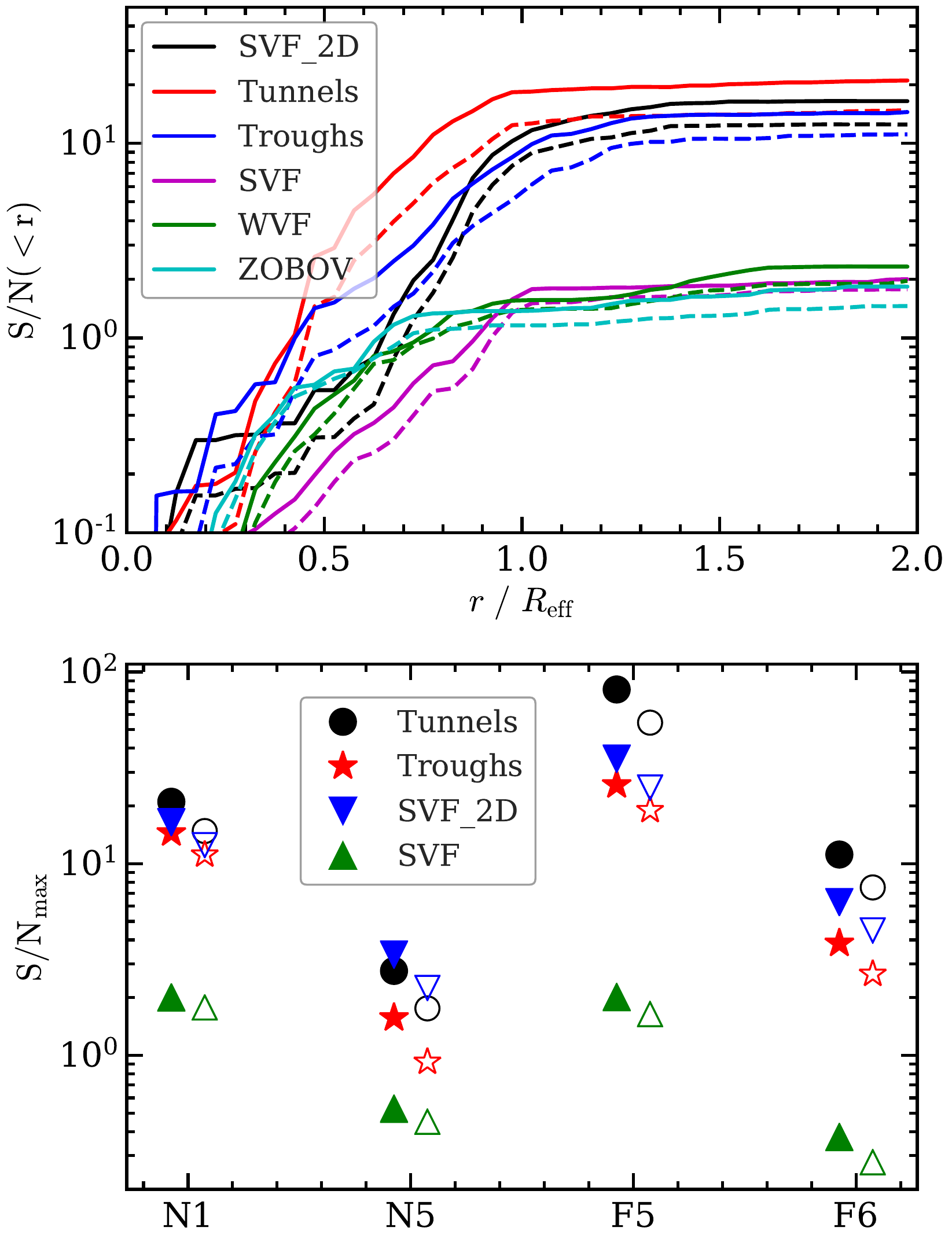}
    \vskip -.2cm
    \caption{Upper panel: The cumulative signal-to-noise of the differences in tangential shear between N1 and GR, for an \LSST{} (solid) and a \Euclid{}-like (dashed) survey. The various colours correspond to different void finding algorithms. Lower panel: The maximum signal-to-noise, S/N$_{\rm{max}}$, of the differences in tangential shear between modified gravity models (nDGP and $f(R)$) and GR. The various symbols show results obtained by different void finders. Filled and open symbols show predictions for \LSST{}- and \Euclid{}-like surveys, respectively. 
    }
    \label{fig:signal_to_noise}
\end{figure}

Large-scale galaxy surveys can be used to measure the weak lensing signal of voids (e.g. \citealt{Clampitt2015,Sanchez2017}). Particularly promising is the measurement of weak lensing by voids with forthcoming wide-field surveys, such as \Euclid{} \citep{euclid} and \LSST{} \citep{lsst}, which will have a sky coverage of  $20,000$ and $18,000$ square degrees, respectively. Here we study the discriminating power of these surveys for distinguishing nDGP models from GR, using the void lensing statistics described in the previous section.

To quantify the degree to which nDGP models will be able to be distinguished from the fiducial GR case, we calculate the signal-to-noise, S/N, value for each of our six void catalogues. The cumulative S/N up to a radial bin $k$ is  defined as:
\begin{equation}
\left(\rm{S/N}\right)^2(<k) = \displaystyle \sum_{i\leq k;\; j\leq k} \delta\gamma_t(i) \;\; {\rm cov}^{-1}(i,j) \;\; \delta\gamma_t(j)
     \label{eq:signal-to-noise} \;,
\end{equation}
where $\delta \gamma_t \equiv \gamma_{t\ \rm{nDGP}} - \gamma_{t\ \rm{GR}}$ is the difference between the tangential shear signals in nDGP and GR, and $\rm{cov^{-1}}$ is the inverse of the tangential shear covariance matrix. The S/N values correspond to the number of sigma that the nDGP models can be distinguished from GR. 

To estimate the mean weak lensing tangential shear measurable by surveys like \Euclid{} and \LSST{}, we follow the procedure described in \citet{Cautun2017}. We assume that both \Euclid{} and \LSST{} overlap with spectroscopic surveys in the redshift range, $0.3 < z < 0.7$, which will have galaxy number densities at least as high as that of the \textsc{sdss-cmass} sample. We then calculate an effective critical surface density, $\Sigma_{c; eff}$, for the survey, that depends on the median redshift, $z_{s\ \rm{med}}$, of the source galaxy distribution. Here we adopt $z_{s\ \rm{med}} = 0.8$ and 1.2 for \Euclid{} and \LSST{}, respectively, obtaining $\Sigma_{c; eff} = 6770$ and 3960 $h M_{\rm{\odot}} \rm{pc^{-2}}$ for each survey. 

Then, the tangential shear in the redshift range, $0.3 < z < 0.7$, can be approximated by 
\begin{eqnarray} \label{eq:shear_survey}
\overline{\gamma_t} = \frac{\Delta \Sigma (z_l = 0.5)}{\Sigma_{c; eff}}\ ,
\end{eqnarray}
where $\Delta \Sigma (z_l = 0.5)$ is the differential surface mass density at $z = 0.5$, which is given in Fig. \ref{fig:void_lensing_profile}.

The covariance matrix contains three main sources of errors: void sample variance, the effect of foreground and background large-scale structures, and the shape noise error due to the intrinsic ellipticities of source galaxies. 
To include these effects, we follow the procedure of \citet{Cautun2017}. We account for void sample variance and large-scale structures by generating a mock light-cone mass distribution between the observer and the source galaxy plane. We place each of the five GR realisation in the redshift range $z=0.3$ to $z=0.7$, which corresponds to the lensing volume of interest. To account for uncorrelated large-scale structures, we place another independent GR realisation in the remaining regions of the mock light-cone, that is from $z=0.0$ to $z=0.3$ and from $z=0.7$ to $z=z_s$. We use these mock mass distributions to calculate the tangential shear for each void. To account for shape-noise, we generate a distribution of randomly placed and randomly oriented source galaxies with the same ellipticity, $\sigma_\epsilon$, and number density, $n_s$, as the target survey. For each void, we then calculate the mean source ellipticity using the same radial bins as the tangential shear signal. We adopt $\sigma_\epsilon=0.22$ and $n_s=30$ and 40 galaxies per arcmin$^{2}$ for \Euclid{} and \LSST{}, respectively. We perform this calculation for all the five GR realisations, and, we further split each realisation into 64 sub-regions which are used to generate 100 bootstrap samples. This leads to 500 estimates, which are used for the calculation of the covariance matrix. For a more detailed description, see \citet{Cautun2017}.

The upper panel of Fig.~\ref{fig:signal_to_noise} shows the cumulative S/N of N1 as a function of the distance from the void centre for the six void finders studied here. The solid and dashed lines show S/N values predicted for \LSST{} and \Euclid{}, respectively. In all the cases, the S/N increases steadily until the void radius, after which it remains relatively constant. The 2D voids have the highest S/N values, with \tunnels{} and \SVFtwoD{} achieving a S/N of $\sim20$ and $15$ at $r \sim R_{\rm eff}$, respectively. Slightly larger S/N values are predicted for \LSST{} {than} \Euclid{}, since the source galaxies from LSST are going to have a higher median redshift, which decreases the effective critical surface density of the survey (see Eq.~19 in \citealt{Cautun2017}), as well as 25 per cent higher galaxy number density, which helps to reduce the shape noise. The 3D finders produce similar S/N values among them, reaching a S/N $\sim 2$.

In the lower panel of Fig.~\ref{fig:signal_to_noise} we compare the highest nDGP S/N values predicted for \LSST{} and \Euclid{}, shown by the filled and open symbols, respectively. We also show results obtained by \citet{Cautun2017} for $f(R)$ gravity models. The horizontal axis categorises the different cosmological models, while the vertical axis show the highest S/N achieved by that model, i.e., the cumulative S/N at $r = 2r_{\rm{void}}$. For the 3D finders, we only show results for the \SVF{}, as all 3D finders produce similar S/N values. 
We find that 2D voids have the largest power to discriminate from GR the two modified gravity theories studied here. While for $f(R)$ models the tunnels are the best choice, for nDGP models the tunnels and \SVFtwoD{} have roughly equal S/N. The 3D voids perform poorly, with \SVF{} having a S/N more than an order of magnitude lower than tunnels. Among the four modified gravity models shown in the figure, N5 is the most difficult one to constrain, with \LSST{} \SVFtwoD{} reaching a S/N of only 3. This panel also shows that void lensing is better in constraining $f(R)$ models than nDGP ones. This can be seen when comparing N1 with F5, and N5 with F6, since each pair of models, e.g. N1 and F5, lead to the same $\sigma_8$ value at $z=0$.

\section{Conclusions} \label{sec:conclusions} 

We have studied cosmic voids in nDGP, a model that is representative of a class of modified gravity theories in which the Vainstein screening hides differences with respect to GR near massive structures. This screening is less efficient in underdense regions, which makes voids promising for constraining these and other related modified gravity theories in cosmology.

We use $N$-body simulations of GR and nDGP universes that share the same initial conditions. We populate dark matter haloes with galaxies using the HOD method. For the GR simulation we use the HOD parameters from \citet{Manera2013}, so that the resulting mock catalogue reproduces the projected two-point clustering and number densities of the \textsc{boss} \textsc{cmass} \textsc{dr9} galaxy sample. The HOD parameters of the nDGP catalogues were tuned to match the projected two-point clustering and number densities of the GR one.

Voids are identified on the mock catalogues using six different void finding algorithms, three of which use the 3D galaxy distribution, and three others use the projected (2D) galaxy distribution. We measure the matter density profile, radial force profile and weak lensing tangential shear profile around void centres. We make predictions for the constraining power of these voids to disentangle nDGP gravity models in the upcoming \LSST{} and \Euclid{} surveys, and we contrast our results with those found in \citet{Cautun2017} for $f(R)$ gravity models.

We first study the average behaviour of voids by stacking them with respect to their centres, the so called spherical stacking procedure. The interiors of voids in nDGP models are more underdense than in GR, with the evacuated mass being piled up at the void edges (see Fig. \ref{fig:void_den_profile}). This is due to the presence in nDGP of a long-range fifth force which points outward from the void centres and thus enhances the evacuation of matter from voids (see Fig.~\ref{fig:void_force_profile}). On average, the ratio of the radial fifth to Newtonian forces is independent of distance and at $z=0.5$ is roughly $10$ and $3$ per cent for the N1 and N5 models, respectively. The underdense nature of nDGP voids leads to a larger void tangential shear signal in nDGP compared to GR. The relative enhancement in tangential shear is around $5$ and $1$ per cent for the N1 and N5 models, respectively, and it is roughly the same for all void finders (see Fig. \ref{fig:void_lensing_profile}). However, 2D void finders such as \SVFtwoD{}, tunnels and troughs have a tangential shear signal an order of magnitude larger than 3D voids, and thus are the most promising void finders for testing departures from GR. After taking into account all major  sources of error (such as the void sample variance, uncorrelated line-of-sight large-scale structures and source galaxy shape noise), we find that \SVFtwoD{} voids and tunnels in the upcoming \LSST{} survey can discriminate N1 and N5 from GR with a signal-to-noise of ${\sim}20$ and ${\sim}3$, respectively (see Fig.~\ref{fig:signal_to_noise}).

The phenomenology of voids in nDGP theories is especially clear when accounting for the fact that voids are highly non-spherical. This can be incorporated in the analysis by stacking voids with respect to their edge. We have implemented this stacking procedure for \WVF{} voids to find that the void interiors show a roughly constant difference in their density profiles between nDGP and GR models. The evacuated mass is deposited within ${\sim}3~h^{-1}\rm{Mpc}$ of the void edges and leads to a pronounced increase in the density of the void boundary in nDGP compared with GR (see Fig. \ref{fig:void_den_profile_boundary}). Calculating stacked tangential shear profiles of 3D voids with respect to their edges leads to roughly a factor of two times larger void weak lensing signal (see Fig. \ref{fig:void_lensing_profile_boundary}), which, while larger than the signal resulting from spherical stacking of 3D voids, it is still lower than the 2D void weak lensing signal.

We have also compared the nDGP void statistics with those of voids in $f(R)$ gravity. Similar to nDGP, $f(R)$ voids are emptier than their GR counterparts. However, we also have found clear differences between the nDGP and $f(R)$ gravities which relate to the different screening mechanisms acting in those theories: Vainshtein versus chameleon screening. By comparing nDGP and $f(R)$ models that lead to similar $\sigma_8$ values, we have found that 3D voids are emptier in nDGP; however the opposite is true for \tunnels{} (see Fig.~\ref{fig:void_den_profile_nDGP_fR}). This is a consequence of the different halo mass functions in nDGP and $f(R)$ models, and the way the \tunnels{} are defined. Fig.~\ref{fig:HMF} shows that haloes less massive than $\sim 10 ^{14}~h^{-1}M_{\rm{\odot}}$ grow faster in $f(R)$ gravity than their counterparts in nDGP. Since they live in low-density environments, these low-mass haloes in chameleon models are mostly unscreened and the fifth force boosts their mass accretion rate. This does not happen as often in nDGP, where even low-mass haloes can be self-screened due to the Vainsthein mechanism. Tunnel boundaries
are populated by typical galaxies which, for a galaxy catalogue with number density, $n_g\sim3\times10^{-4}~(h^{-1}{\rm Mpc})^{-3}$,  
are located in haloes with average masses ${\sim}10 ^{13}~h^{-1}M_{\rm{\odot}}$. Since their host haloes are more massive in $f(R)$ gravity than in nDGP, it results in tunnels having a more prominent ridge and more underdense interiors in the former model. This leads to \tunnels{}, and 2D void finders in general, to have a much larger signal-to-noise for testing $f(R)$ gravities than nDGP models with similar $\sigma_8$ (see Fig. \ref{fig:signal_to_noise}).

We have found major differences between the void fifth force profiles in $f(R)$ gravity versus in nDGP from our simulations. The amplitudes of the fifth force profile in $f(R)$ clearly decreases with the increase of the size of voids, but they remain similar around nDGP voids of different sizes. This is consistent with expectations for the chameleon screening and the Vainshtein screening mechanisms associated with these two models, and help to explain the difference of void profiles between these two models. The unique size-dependence of the fifth force around voids in the $f(R)$ model offers possibilities for it to be distinguished from other models via dynamical approaches.

Both nDGP and $f(R)$ gravities produce larger void lensing signals than GR, with the strength of the deviation from GR depending on parameters of the two modified gravity theories. This means that if one were to measure a deviation from the GR prediction of void lensing, it would be difficult to tell to which of the two modified gravity theories it corresponds to. However, we have found that by comparing the lensing signal from two different void finding algorithms, it may be possible to disentangle nDGP from $f(R)$ gravity (Fig.~\ref{fig:lensing_nDGP_vs_fR}). The 2D void finders are promising in this regard, the combination of \tunnels{} and \troughs{} in particular.

Overall, we find that cosmic voids offer excellent environments for constraining nDGP-like gravity theories, and also for improving our understanding of the Vainsthein screening mechanism present in similar theories. 2D void finders will play a critical role for trying to constrain modified gravity in future large-scale galaxy surveys, especially as tools to differentiate between two alternative classes of modified gravity theories, such as nDGP and $f(R)$ gravity. Another probe that has shown great potential when studying deviations from GR in these models are the velocity profiles around voids (\citealt{Cai2014, Falck2017}). In the next paper of this series, we aim to study the constraining power of redshift-space distortions around voids in $f(R)$ gravity and nDGP, thus widening the set of void observables that can be used as tests for alternative models to $\Lambda \rm{CDM}$.

\section*{Acknowledgements}

We thank the anonymous referee for their insightful comments.
EP is supported by CONICYT-PCHA/Doctorado Nacional (2017-21170093) and acknowledges support from CONICYT project Basal AFB-170002. MC was supported by Science and Technology Facilities Council (STFC) [ST/P000541/1]. SB is supported by Harvard University through the ITC Fellowship. YC was supported by the European Research Council under grant numbers 670193. NP received support from Fondecyt Regular 1150300, BASAL CATA PFB-06, Anillo ACT-1417. BL is supported by an European Research Council Starting Grant (ERC-StG-716532-PUNCA) and STFC (ST/L00075X/1, ST/P000541/1). 
This project has received funding from the European Union's Horizon 2020 Research and Innovation Programme under the Marie Sk\l odowska-Curie grant agreement No 734374.     
This work used the DiRAC Data Centric system at Durham University, operated by the Institute for Computational Cosmology on behalf of the STFC DiRAC HPC Facility (www.dirac.ac.uk). This equipment was funded by BIS National E-infrastructure capital grant ST/K00042X/1, STFC capital grants ST/H008519/1 and ST/K00087X/1, STFC DiRAC Operations grant ST/K003267/1 and Durham University. DiRAC is part of the National E-Infrastructure.
Part of the analysis was done on the Geryon cluster at the Centre for Astro-Engineering UC. The Anillo ACT-86, FONDEQUIP AIC-57, and QUIMAL 130008 provided funding for several improvements to the Geryon cluster.




\bibliographystyle{mnras}
\bibliography{mnras_template} 



\appendix


\bsp	
\label{lastpage}
\end{document}